\documentclass[11pt,english]{article}
\usepackage[T1]{fontenc}
\usepackage[latin9]{inputenc}
\usepackage{geometry}
\usepackage{xcolor}
\usepackage{pdfcolmk}
\usepackage{babel}
\usepackage{rotating}
\usepackage{mathrsfs}
\usepackage{url}
\usepackage{multirow}
\usepackage{amsmath}
\usepackage{amsthm}
\usepackage{amssymb}
\usepackage{graphicx}
\usepackage{setspace}
\usepackage[authoryear]{natbib}
\PassOptionsToPackage{normalem}{ulem}
\usepackage{ulem}
\usepackage[unicode=true,pdfusetitle,
 bookmarks=true,bookmarksnumbered=false,bookmarksopen=false,
 breaklinks=false,pdfborder={0 0 0},pdfborderstyle={},backref=false,colorlinks=false]
 {hyperref}

\makeatletter

\providecommand{\tabularnewline}{\\}
\providecolor{lyxadded}{rgb}{0,0,1}
\providecolor{lyxdeleted}{rgb}{1,0,0}
\DeclareRobustCommand{\lyxsout}[1]{\ifx\\#1\else\sout{#1}\fi}

\theoremstyle{remark}
\newtheorem{rem}{\protect\remarkname}
\theoremstyle{plain}
\newtheorem{assumption}{\protect\assumptionname}
\theoremstyle{plain}
\newtheorem{lem}{\protect\lemmaname}
\theoremstyle{plain}
\newtheorem{thm}{\protect\theoremname}
\theoremstyle{definition}
 \newtheorem{example}{\protect\examplename}
\theoremstyle{definition}
\newtheorem*{example*}{\protect\examplename}
\theoremstyle{plain}
\newtheorem{cor}{\protect\corollaryname}
\theoremstyle{remark}
\newtheorem*{rem*}{\protect\remarkname}

\usepackage[auth-lg]{authblk}

\theoremstyle{definition}

\newtheoremstyle{proof}
  {\topsep}   % ABOVESPACE
  {\topsep}   % BELOWSPACE
  {}  % BODYFONT
  {0pt}       % INDENT (empty value is the same as 0pt)
  {} % HEADFONT
  {}         % HEADPUNCT
  {0pt} % HEADSPACE
  {}          % CUSTOM-HEAD-SPEC

\theoremstyle{proof}

\let\counterwithin\relax
\usepackage{chngcntr}
\usepackage{apptools}

\AtAppendix{\counterwithin{lem}{section}}

\usepackage{tikz}
\usetikzlibrary{arrows.meta}
\usetikzlibrary{decorations.pathreplacing}

\@ifundefined{showcaptionsetup}{}{%
 \PassOptionsToPackage{caption=false}{subfig}}
\usepackage{subfig}
\makeatother

\providecommand{\assumptionname}{Assumption}
\providecommand{\corollaryname}{Corollary}
\providecommand{\examplename}{Example}
\providecommand{\lemmaname}{Lemma}
\providecommand{\remarkname}{Remark}
\providecommand{\theoremname}{Theorem}

\begin{document}

\title{Forward-Selected  Panel Data Approach\\for Program Evaluation}
\author[a]{
	Zhentao Shi\thanks{
	Corresponding author. Email addresses: \url{zhentao.shi@cuhk.edu.hk} (Z.~Shi), \url{jingyi.huang@barclays.com} (J.~Huang).}
}
\author[b]{Jingyi Huang}
\affil[a]{Department of Economics, the Chinese University
of Hong Kong, Sha Tin, New Territories, Hong Kong SAR, China}
\affil[b]{Barclays Capital
Asia Ltd., Cheung Kong Center, 2 Queen\textquoteright s Road Central,
Hong Kong, Hong Kong SAR, China}
\date{}

\renewcommand\Affilfont{\small \itshape}
\renewcommand\Authfont{\large}

\maketitle

\begin{abstract}
Policy evaluation is central to economic data analysis, but economists
mostly work with observational data in view of limited opportunities
to carry out controlled experiments. In the potential outcome framework,
the panel data approach (Hsiao, Ching and Wan, 2012) constructs the
counterfactual by exploiting the correlation between cross-sectional
units in panel data. The choice of cross-sectional control units,
a key step in its implementation, is nevertheless unresolved in data-rich
environment when many possible controls are at the researcher's disposal.
We propose the forward selection method to choose control units, and
establish validity of the post-selection inference. Our asymptotic
framework allows the number of possible controls to grow much faster
than the time dimension. The easy-to-implement algorithms and their
theoretical guarantee extend the panel data approach to big data settings.
\end{abstract}
\vspace{1cm}



\newpage{}

\normalsize
\renewcommand{\hat}{\widehat}

\section{Introduction}

A controlled experiment compares outcomes of a treatment group with
those from a control group. It is the golden standard for scientific
research. While the randomized controlled trials are useful in understanding
economic mechanisms \citep{duflo2007using,banerjee2009experimental},
for large-scale questions economists mostly have access to observational
datasets only. For example, in economic research we rarely enjoy the
luxury to implement a controlled experiment at a national level that
would affect millions of people\textemdash such an exercise can be
prohibitively expensive or ethically unacceptable. Instead, economists
resort to constructing counterfactuals from observational data for
policy evaluation. A counterfactual is the potential outcome that
can be perceived but has never happened in the real world.

In view of the lack of genuine control groups in many important economic
empirical questions, \citet{hsiao2012panel} (HCW, henceforth) propose
the panel data approach (PDA) to exploit the correlation between cross-sectional
units in estimating the counterfactual. Consider, for simplicity,
that a single treatment (\emph{treatment} and \emph{intervention}
are used as synonyms) is carried out during the observed time window.
PDA is a linear regression on the cross-sectional units in the pre-treatment
data, and then these estimated coefficients are used to extrapolate
the counterfactual of no policy intervention in the post-treatment
period. Its convenience attracts many applications and extensions,
for example \citet{bai2014property}, \citet{fujiki2015disentangling},
\citet{ouyang2015treatment}, \citet{ke2017china}, to name a few.
Compared with the popular difference-in-difference method, the combination
of control units allows time-varying treatment effect. Alternatively,
\citet{abadie2003economic} and \citet{abadie2010synthetic} advocate
the \emph{synthetic control method} (SCM). \citet{gardeazabal2017empirical}
and \citet{wan2018panel} compare PDA and SCM in simulations and empirical
applications.

Choices of the control units directly affect PDA's estimation and
inference results, and thus a systematic variable selection scheme
is of vital practical importance. HCW experiment with the Akaike information
criterion (AIC) and the corrected AIC (AICC), and \citet{du2015home}
recommend the latter for consistent variable selection. These conventional
variable selection methods compute an information criterion for each
candidate model to identify the ``best subset''. In PDA the total
number of candidate models is $2^{N}$, where $N$ is the number of
available potential control units. In spite of the state-of-the-art
computing technology, exhaustive search quickly becomes too time-consuming
for a moderate $N$. The exhaustive enumeration is inapplicable in
the era of big data when data-rich environments offer information
at an unprecedented scale.

Furthermore, besides the computational
difficulty, a large cross-sectional dimension also challenges PDA's
theoretical justification. As PDA is often applied to aggregate data
with low-frequency temporal observations, HCW's conventional ``fixed
$N$, large $T$''\footnote{Here $T$ represents the size of the time dimension in a generic panel
data setting. We will elaborate it when introducing the model.} asymptotic framework is unlikely to deliver satisfactory approximation
in empirical studies where $N$ is comparable to $T$, or even exceeds
$T$. To overcome the high dimensionality, \citet{li2016estimation}
suggest using Lasso \citep{tibshirani1996regression} but provide
no theoretical foundation. \citet{carvalho2016arco} develop the Lasso
theory under a general framework called Artificial Counterfactual
(ArCo).

This paper studies estimation and inference of the average treatment
effect (ATE) by PDA when a large number of candidate cross-sectional
control units are present. Motivated by real data applications, we
tackle variable selection in high dimension.

\begin{example}
HCW's original empirical application 
evaluates the effect of a trade treaty on Hong Kong's GDP growth rate.
Their dataset contains $61$ quarterly observations, and $24$ economies serve as control units.
Hong Kong is an Asian trade and financial hub doing business with many partners, however. 
Given nearly 200 countries and territories in the world, how to deal with a much larger pool of control units?
\end{example}

\begin{example}
In Section \ref{sec:Empirical-Applications} we
intend to quantify the impact of China's anti-corruption campaign
on the luxury watch import.
The time series of six years' monthly luxury watch import is collected from 
a comprehensive United Nations dataset, which also provides the amounts of 88 categories of commodities that China imports.
Without knowing which commodities are associated with the import of watches, 
we take an agnostic view and employ the universe of control units.
\end{example}

This paper contributes, toward the growing toolkit of program evaluation, 
a user-friendly procedure with asymptotic guarantee
to make it possible to incorporate all potential control units with accessible time series.
It alleviates the arbitrariness of control units selection --- an essential step in program evaluation. 

In terms of the algorithms, we
suggest using forward selection to choose a sequence of control units
one by one up to a desirable number. Involving only a series of OLS
regressions, forward selection is computationally efficient. For hypothesis
testing about the ATE, we advocate calculating a conventional $t$-statistic
conditioning on the control variables chosen by forward selection
and then comparing it with a quantile of the standard normal distribution.
We call this two-step method the \emph{forward-selected PDA} (fsPDA),
as in the title of the paper.

Despite the simplicity of the algorithms, the underlying asymptotic theory for fsPDA nevertheless
demands careful development and justification. The environment of
independently and identically distributed (i.i.d.)~data in which
most high-dimensional statistical problems are investigated is too
restrictive for economic questions with temporal observations. Accommodating
heterogeneous weakly dependent time series, we establish our theory
in the asymptotic framework allowing $N/T\to\infty$ when both dimensions
diverge.

A unique innovation is that our theory is valid in regression models
no matter whether the ``true'' underlying coefficients are dense
or sparse. It differentiates us from the vast literature of high-dimensional
statistics that counts on sparse regression models for asymptotic
results. Here \emph{dense} models impose no restrictions on the coefficients
\textemdash{} in principle all coefficients can be simultaneously
non-zero.\footnote{Dense model are of rising interest in economics where observed variables
are interconnected and little prior knowledge ensures that the driven
forces fall into a few of key variables \citep{giannone2017economic}.} In contrast, \emph{sparsity} means most of the regression coefficients
are exactly zero or too small to matter. As to be discussed in Section
\ref{subsec:PDA_Model-and-Hypo}, PDA is motivated from a factor model
which induces a dense regression in general. We make it clear that
the inference of ATE in the post-treatment data for correct test size
does not require consistent estimation of the true underlying high-dimensional
coefficients in the data generating process (DGP). Instead, it is
sufficient if we can recover the linear projection coefficients associated
with a small subset of control variables.

We add to the literature of both variable selection and statistical
inference in the dense model setting. (i) We show that forward selection
is capable of reducing the variance of the regression error as much
as that of a computationally infeasible best subset. Although forward
selection is by no means a new algorithm, in the past its theory is
established either in sparse statistical models \citep{Zhu2017forward}
or dense population models \citep{das2011submodular}. We are unaware
of its theory in dense models with sampling error. (ii) Many asymptotic
normality results in high-dimensional models are pointwise asymptotics
under a single DGP, for example the so-called \emph{oracle property}
\citep{fan2001variable,zou2006adaptive}. Our inferential theory is
uniform under DGPs that satisfy a set of conditions. In other words,
the seemingly naive practice of conventional normal inference is valid
and the randomness stemming from the step of variable selection can
be safely ignored. This validity comes from the special structure
of PDA, in which the pre-treatment and post-treatment periods are
naturally separated into two disjoint segments. The control units
are selected from the pre-treatment data only, and under weak dependence
they become asymptotically independent of the post-treatment data
on which the test statistic is based. 

This paper fits in the theme of research to better connect settings
of economic interest with modern high-dimensional statistics. It transpires
that the economic context of PDA underpins our unsophisticated procedure
and circumvents statistical challenges encountered by post-selection
uniform inference in a single dataset. The PDA environment also helps
with other technical building blocks. For instance, the restricted
eigenvalue condition \citep{bickel2009simultaneous} is often viewed
as a necessary but somewhat \emph{ad hoc} assumption in high-dimensional
regressions. Rather than imposing it, we argue that a version of the
restricted eigenvalue condition is a natural implication of the underlying
latent factor model that motivates PDA. 
The theoretical properties that we established for PDA are supported
by extensive Monte Carlo simulations. 

\medskip{}

\textbf{Literature Review. }Various greedy variable selection algorithms
have been studied in operational research, statistics and econometrics.
Working with random samples, \citet{wang2009forward}, \citet{Zhu2017forward},
and \citet{Zhu2019JASA} analyze forward selection as a device for
model determination in statistical ultrahigh-dimensional sparse regressions.
\citet{kozbur2017testing}, \citet{kozbur2018sharp} and \citet{hansen2018targeted}
investigate test-based stopping criteria and post-selection inference.
Our paper extends the operational research by \citet{das2011submodular}
and \citet{das2018approximate} who highlight the key role played
by \emph{submodular ratio} in the population model analysis of forward
selection. When we carry forward selection over into panel data, we
must cope with sampling uncertainty as well as temporal dependence.
The greedy nature of forward selection is closely related to the component-wise
boosting \citep{buhlmann2006,luo2016high} which is familiar to econometricians
\citep{bai2009boosting,shi2015REL,fonseca2018boost,Phillips2019boosting}.
Alternatively, \citet{carvalho2016arco} studies asymptotic validity
of ATE estimation by Lasso-type methods in sparse models.

Uniform inference after variable selection is a difficult statistical
issue, as pointed out by \citet{leeb2005model} and \citet{leeb2006can}.
Proposed solutions usually resort to non-standard methods, for example
\citet{berk2013valid}, \citet{fithian2014optimal} and \citet{tibshirani2018uniform},
when model selection and testing are carried out within a single dataset.
Predictive inference, however, provides an amenable environment to
work with and \citet{leeb2009conditional} shows post-selection asymptotic
normality. Another related line of uniform inference literature tries
to correct the shrinkage bias in high-dimensional regressions \citep{belloni2014uniform,belloni2017program,javanmard2018debiasing}.
In our paper we always use OLS to estimate coefficients so we are
free from shrinkage biases caused by penalized estimation.

\medskip{}

\textbf{Notations.} Unless explicitly defined otherwise, we use a
plain letter, say ``$x$'', to denote a scalar, a boldface lowercase
letter ``$\mathbf{x}$'' to denote a column vector, and a boldface
uppercase letter ``$\mathbf{X}$'' to denote a matrix. The square
matrix $\mathbf{I}$ is an identity matrix. \textbf{1}$\left\{ \cdot\right\} $
is the indicator function. For a real number, $\Phi(\cdot)$ is the
cumulative distribution function of the standard normal distribution
$N\left(0,1\right)$, $\left\lceil \cdot\right\rceil $ is the ceiling
function and $\left\lfloor \cdot\right\rfloor $ is the floor function.
For a square matrix, $\left(\cdot\right)^{-}$ is the Moore-Penrose
generalized inverse, and $\phi_{\min}\left(\cdot\right)$ and $\phi_{\max}\left(\cdot\right)$
are the minimum eigenvalue and the maximum eigenvalue, respectively.
The cardinality of a discrete set $U$ is denoted as $\left|U\right|$.
A vector with a discrete set as its subscript, $\mathbf{x}_{U}:=(x_{j})_{j\in U}$,
makes a $\left|U\right|$-element subvector of $\mathbf{x}$. $\left\Vert \cdot\right\Vert _{2}$
and $\left\Vert \cdot\right\Vert _{1}$ are the usual $L_{2}$ and
$L_{1}$ vector norms, respectively.

Now we introduce PDA's panel data setting. $(N+1)$ cross-sectional
units in a panel data are indexed by $\mathcal{N}_{0}:=\left\{ 0,1,\ldots,N\right\} $,
in which $j=0$ indexes the sole \emph{treated unit} whereas $\mathcal{N}:=\left\{ 1,\ldots,N\right\} $
is the index set of the $N$ \emph{control units}. In the potential
outcome framework, let $y_{jt}^{1}$ and $y_{jt}^{0}$ be the outcomes
of the unit $j$ at time $t$ with and without a policy intervention,
respectively. We cannot witness $y_{jt}^{1}$ and $y_{jt}^{0}$ simultaneously;
instead we observe $y_{jt}=y_{jt}^{0}(1-d_{jt})+y_{jt}^{1}d_{jt}$,
where $d_{jt}$ is a dummy variable equal to 1 if the $j$-th unit
is under intervention at time $t$; otherwise $d_{jt}=0$.

\begin{figure}[h]
\centering

\begin{tikzpicture}

\draw (-6,0) -- (-4,0) ; %edit here for the axis
\draw[dashed] (-4,0) -- (-2,0) ; %edit here for the axis
\draw (-2,0) -- (2,0) ; %edit here for the axis
\draw[dashed] (2,0) -- (4,0) ; %edit here for the axis
\draw[-stealth] (4,0) -- (6,0) ; %edit here for the axis

\foreach \x in  {-5, -4,-2,-1,0 1,2,4, 5} % edit here for the vertical lines
\draw[shift={(\x,0)},color=black] (0pt,3pt) -- (0pt,-3pt);
\foreach \x in {-2,-1, 0, 1,2} % edit here for the numbers
\draw[shift={(\x,0)},color=black] (0pt,3pt) -- (0pt,-3pt) node[below] 
{$\x$};
\draw[shift={(-4,0)},color=black] (0pt,3pt) -- (0pt,-3pt) node[below] 
{$ -T_1 $};
\draw[shift={(4,0)},color=black] (0pt,3pt) -- (0pt,-3pt) node[below] 
{$ T_2 $};
\draw[<-] (0,0.3)--++(90:0.5) node[above]{Treatment};

\draw (6, 0.1) node[above] {$ \infty $};
% \draw (5, 0.1) node[right] {time};
\draw (-6, 0.1) node[above] {$ -\infty $};
\draw [decorate,
          decoration={brace,amplitude=4pt,      },xshift=0pt,yshift=+10pt]
          (1,0) -- (4,0) 
          node [black,midway,yshift=+0.5cm]
          {$ \mathcal{T}_2 $};
\draw [decorate,
          decoration={brace,amplitude=4pt,      },xshift=0pt,yshift=+10pt]
          (-4,0) -- (-1,0) 
          node [black,midway,yshift=+0.5cm]
          {$ \mathcal{T}_1 $};
        
\end{tikzpicture}

\caption{\label{fig:timeline} Timeline of the Times Series, Observations,
and Treatment}
\end{figure}
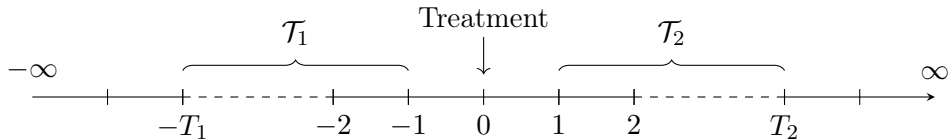

The time dimension of the panel data is shown in the timeline diagram
in Figure \ref{fig:timeline}. The time series extends from $-\infty$
to $\infty$, while we econometricians observe a time window $\left\{ -T_{1},\ldots-1,0,1,\ldots T_{2}\right\} $
for some $T_{1},T_{2}\in\mathbb{N}$. Without loss of generality,
a policy intervention occurs at time $t=0$, which partitions the
observed interval into two sections: a pre-treatment period $\mathcal{T}_{1}:=\left\{ -T_{1},\ldots,-1\right\} $
and a post-treatment period $\mathcal{T}_{2}:=\left\{ 1,\ldots,T_{2}\right\} $,
with lengths $T_{1}=\left|\mathcal{T}_{1}\right|$ and $T_{2}=\left|\mathcal{T}_{2}\right|$.
Denote $\mathcal{T}:=\mathcal{T}_{1}\cup\mathcal{T}_{2}$, and $T:=\left|\mathcal{T}\right|$.

The mathematical expectation of a generic random variable $x_{t}$
is denoted as $E\left[x_{t}\right]$. For a heterogeneous time series,
we define $\mathcal{E}_{\left(1\right)}\left[x_{t}\right]:=T_{1}^{-1}\sum_{t\in\mathcal{T}_{1}}E\left[x_{t}\right]$
as the average of the expectations $(E[x_{t}])_{t\in\mathcal{T}_{1}}$
in the pre-treatment period, and similarly $\mathcal{E}_{\left(2\right)}\left[x_{t}\right]:=T_{2}^{-1}\sum_{t\in\mathcal{T}_{2}}E\left[x_{t}\right]$
as the average of the expectations after the treatment. We define
$\mathbb{E}_{\left(1\right)}\left[x_{t}\right]:=T_{1}^{-1}\sum_{t\in\mathcal{T}_{1}}x_{t}$
as the pre-treatment sample mean, and $\mathbb{E}_{\left(2\right)}\left[x_{t}\right]=T_{2}^{-1}\sum_{t\in\mathcal{T}_{2}}x_{t}$
as the post-treatment sample mean.

\medskip{}

\textbf{Plan.} The rest of the paper is organized as follows. Section
\ref{sec:Panel-Data-Approach} introduces PDA and describes forward
selection and the post-selection ATE inference. Section \ref{sec:Asymptotic-Analysis}
presents asymptotic analysis of fsPDA. Section \ref{sec:Simulations}
reports the simulation results, and Section \ref{sec:Empirical-Applications}
carries out an empirical example. All proofs are provided in the appendix.
Moreover, an online supplement is prepared with further comparison
of methods, one more empirical example, and additional simulation
results.
Replication data and a documented R package {\tt fsPDA} are hosted at 
\url{https://github.com/zhentaoshi/fsPDA}.

\section{\label{sec:Panel-Data-Approach} Panel Data Approach in High Dimension}

\subsection{Model and Hypothesis \label{subsec:PDA_Model-and-Hypo}}

PDA is motivated from a factor model. For the completeness of the
paper, we briefly summarize HCW's proposal. Consider a pure factor
model in which all cross-sectional units share at most $K$ common
factors:
\begin{equation}
y_{jt}^{0}=\boldsymbol{\lambda}_{j}'\mathbf{f}_{t}+e_{jt},\ \ j\in\mathcal{N}_{0},\ t\in\mathcal{T},\label{eq:factor_single}
\end{equation}
 where $\mathbf{f}_{t}$ is a mean zero $K$-vector of latent factors,
$\boldsymbol{\lambda}_{j}$ is a $K$-vector of factor loading, and
$e_{jt}$ is a mean zero idiosyncratic error component orthogonal
to the factors.\textbf{}\footnote{For simplicity of presentation, we assume $E[y_{jt}^{0}]=0$ for all
$j\in\mathcal{N}_{0},\ t\in\mathcal{T}$ and the linear regressions
in Section \ref{sec:Panel-Data-Approach} does not include an intercept.
While the presence of the intercept incurs extra notation, incorporating
it does not affect the asymptotic theory \citep[p.104]{buhlmann2011statistics}.
The intercept will be included in the empirical applications.} 

HCW assume only one unit ($j=0$) is exposed to a policy intervention,
so the intervention does not affect the outcomes of all other units
$i\in\mathcal{N}$. After the treatment, the observed outcome for
the treated unit is
\begin{equation}
y_{0t}^{1}=y_{0t}^{0}+\Delta_{t},\ \ t\in\mathcal{T}_{2}\label{eq:Delta}
\end{equation}
where $\Delta_{t}$ is the treatment effect at time $t$. PDA is interested
in the null hypothesis of, without loss of generality, zero ATE
\[
\mathbb{H}_{0}:\ \mathcal{E}_{\left(2\right)}\left[\Delta_{t}\right]=0.
\]
 For only $y_{0t}^{1}$ is observable after the intervention, to evaluate
the treatment effect we must estimate the counterfactual $y_{0t}^{0}$
for $t\in\mathcal{T}_{2}$ from the observed data. \citet{li2016estimation}
show that, based on the factor model, there exists an $N$-vector
$\boldsymbol{\boldsymbol{\beta}}_{\mathcal{N}}^{0}$ such that the
outcome of the treated unit can be written as a linear combination
of the outcomes of the control units plus an orthogonal error 
\begin{equation}
y_{0t}^{0}=\mathbf{y}_{\mathcal{N}t}'\boldsymbol{\boldsymbol{\beta}}_{\mathcal{N}}^{0}+\varepsilon_{t},\mbox{ }\mbox{for}\mbox{ }t\in\mathcal{T}\mbox{,}\label{eq:y0}
\end{equation}
where $\mathbf{y}_{\mathcal{N}t}:=(y_{jt})_{j\in\mathcal{N}}$ is
an $N$-dimensional vector. Although the regression equation (\ref{eq:y0})
\textemdash{} PDA's workhorse for estimation and inference \textemdash{}
is derived from the factor model (\ref{eq:factor_single}), the factor
model itself merely serves as a motivation but is irrelevant to PDA's
implementation. 

With the pre-treatment sub-sample $\mathcal{T}_{1}$, HCW advocate
estimating $\hat{\boldsymbol{\beta}}_{\mathcal{N}}$ by ordinary least
squares (OLS) or generalized least squares (GLS), and predicting the
counterfactual as $\hat{y}_{0t}^{0}=\mathbf{y}_{\mathcal{N}t}'\hat{\boldsymbol{\beta}}_{\mathcal{N}}$
for $t\in\mathcal{T}_{2}$ and the treatment effect $\hat{\Delta}_{t}=y_{0t}^{1}-\hat{y}_{0t}^{0}$.
The inference is routine. They construct the usual $t$-statistic
based on the sample mean of $\{\hat{\Delta}_{t}\}_{t\in\mathcal{T}_{2}}$
and its long-run variance, and then compare the absolute value of
the $t$-statistic with some quantile of $N\left(0,1\right)$, say
1.96 for two-sided test of size 5\%, to decide whether the null should
be rejected. 

\subsection{\label{subsec:algorithm-FS} Forward Selection}

The estimate of PDA depends on the choice of the control units. When
the number of potential controls is large, HCW's information criterion
approach will encounter computational difficulty in exhaustive search.
To solve this problem, we propose using the forward selection method
\citep[Chapter 3.3.3]{hastie2009bible}. In the first iteration, we
regress $\left(y_{0t}\right)_{t\in\mathcal{T}_{1}}$ on $\left(y_{jt}\right)_{t\in\mathcal{T}_{1}}$
for each $j\in\mathcal{N}$, and choose the one that maximizes the
R-squared $\mathscr{R}^{2}\left(\left\{ j\right\} \right)$ of OLS,
where $\left\{ j\right\} $ in the parenthesis stresses the set of
control units on which the R-squared $\mathscr{R}^{2}$ is based.
We denote the index of the maximizer as $\hat{j}_{1}$ and let $\hat{U}_{1}=\{\hat{j}_{1}\}$
be a single-element set. In the $r$-th iteration, where $r=2,..,R$,
we run OLS of $\left(y_{0t}\right)_{t\in\mathcal{T}_{1}}$ on $(\mathbf{y}_{\hat{U}_{r-1}t})_{t\in\mathcal{T}_{1}}$
together with another single $\left(y_{jt}\right)_{t\in\mathcal{T}_{1}}$
for each $j\in\mathcal{N}\backslash\hat{U}_{r-1}$, choose the one
\textemdash{} denoted as $\widehat{j}_{r}$ \textemdash{} that maximizes
the corresponding R-squared $\mathscr{R}^{2}(\widehat{U}_{r-1}\cup\{j\})$,
and incorporate it into the selected set $\widehat{U}_{r}=\widehat{U}_{r-1}\cup\{\widehat{j}_{r}\}$.
The total number of iterations, $R$, is a tuning parameter specified
by the user. The algorithm is described formally as follows.
\begin{description}
\item [{Step1:}] Choose the number of total iterations $R\in\mathbb{N}$.
Set the initial iteration index as $r=0$ and the selected set as
$\widehat{U}_{0}=\emptyset$.
\item [{Step2.1:}] Update the iteration index $r\leftarrow r+1$; \textbf{Step
2.2:} Get $\hat{j}_{r}=\underset{j\in\mathcal{N}\backslash\hat{U}_{r-1}}{\arg\max}\ \mathscr{R}^{2}(\widehat{U}_{r-1}\cup\{j\})$;
\textbf{Step 2.3:} Update the selected set as $\hat{U}_{r}=\hat{U}_{r-1}\cup\{\hat{j}_{r}\}$.
\item [{Step3:}] Repeat \textbf{Step 2.1-2.3} until $r>R$.
\end{description}
\begin{rem}
When we were preparing this manuscript, \citet[p.467]{hsiao2019panel}
independently experimented a similar algorithm, which they call the
\emph{stepwise regression method}, in Monte Carlo simulations and
empirical applications. They cite Lasso penalty to stop the iteration.
Nevertheless, they provide no theoretical justification for such an
algorithm.
\end{rem}
The above forward selection procedure is a greedy algorithm that takes
the most aggressive direction in each step to increase the R-squared,
or equivalently to reduce the sum of squared residuals, conditional
on the variables that are already included. Once a variable is selected,
there is no mechanism to drop it. Greedy algorithms are popular in
modern machine learning. For example, \citet{breiman2001random} grows
regression trees by splitting a single variable each time at the deepest
descent, and \citet{buhlmann2006}'s componentwise boosting seeks
the most greedy variable without adjusting other coefficients.

\subsection{\label{subsec:algo-Inference}Post-Selection Inference }

The ultimate goal of PDA is statistical inference for the ATE. If
we had prior knowledge about an index set $U\subset\mathcal{N}$ of
relevant control units, we would naturally carry out the following
procedure. We would regress $\left(y_{0t}\right)_{t\in\mathcal{T}_{1}}$
on $(\mathbf{y}_{Ut}=(y_{jt})_{j\in U})_{t\in\mathcal{T}_{1}}$ to
obtain the coefficient $\widehat{\boldsymbol{\beta}}_{U}$ and predict
the counterfactual $\widehat{y}_{0t}^{0}\left(U\right):=\mathbf{y}_{Ut}'\widehat{\boldsymbol{\beta}}_{U}$.
Next, we would estimate the treatment effect based on the set $U$
as 
\[
\hat{\Delta}_{Ut}:=y_{0t}^{1}-\widehat{y}_{0t}^{0}\left(U\right)=y_{0t}^{1}-\mathbf{y}_{Ut}'\widehat{\boldsymbol{\beta}}_{U},\ t\in\mathcal{T}_{2}.
\]
Let $\widehat{\rho}_{\tau U}^{2}:=T_{2}^{-1}\sum_{t,s\in\mathcal{T}_{2}}(\widehat{\Delta}_{Ut}-\bar{\Delta}_{U})(\widehat{\Delta}_{Us}-\bar{\Delta}_{U})\cdot\boldsymbol{1}\left\{ \left|t-s\right|\leq\tau\right\} $
be a heteroskedasticity- and autocorrelation-consistent (HAC) estimator
of the long-run variance, where $\tau$ is the number of lags and
$\bar{\Delta}_{U}:=\mathbb{E}_{\left(2\right)}[\hat{\Delta}_{Ut}]$.
We calculate the $t$-statistic 
\begin{equation}
\mathcal{Z}_{U}:=\widehat{\rho}_{\tau U}^{-1}\cdot\sqrt{T_{2}}\bar{\Delta}_{U},\label{eq:z-test}
\end{equation}
which depends on $\tau$ and $T_{2}$ while we suppress them for conciseness.
We would reject the null hypothesis at size $a$ when $|\mathcal{Z}_{U}|>\Phi^{-1}\left(1-a/2\right)$,
provided that the distribution of $\mathcal{Z}_{U}$ can be approximated
by $N\left(0,1\right)$.

In reality we rarely know in advance a set of relevant control units
$U$. We suggest using $\widehat{U}_{R}$, the set chosen by forward
selection, to substitute the generic $U$ in the $t$-statistic in
the above paragraph. That is, we reject the null hypothesis $\mathbb{H}_{0}$
at 5\% size if $|\mathcal{Z}_{\widehat{U}_{R}}|>1.96$. Although $\widehat{U}_{R}$
is a random set determined by the pre-treatment data, we will show
$N\left(0,1\right)$ is a reasonable approximation to the statistic
$\mathcal{Z}_{\widehat{U}_{R}}$ under the null along with mild assumptions
of weak temporal dependence.

There are two tuning parameters in fsPDA, $R$ for the total
number of selected variables and $\tau$ for the long-run variance
estimation. We suggest using \citet{wang2009shrinkage}'s \emph{modified
Bayesian information criterion} (modified BIC) to choose $R$, while
the choice of $\tau$ has been well studied in econometrics literature
\citep{newey1987simple,andrews1991heteroskedasticity}.

Before we conclude this section, we emphasize that we do not attempt
to directly estimate the factor model due to the following reasons.
(i) In the PDA framework the factor model is an abstraction independent
of the algorithm based on linear regressions, and this regression
approach is also followed by \citet{li2016estimation} and \citet{carvalho2016arco}.
(ii) To conduct inference literally in the factor model, we will need
to estimate the $\left(N+1\right)\times\left(N+1\right)$ covariance
matrix for the idiosyncratic noises $\left(e_{jt}\right)_{j\in\mathcal{N}_{0}}$,
which involves $\left(N+2\right)\left(N+1\right)/2$ unknown entries
so other sparse matrix estimation techniques have to be invoked for
dimension reduction.

\section{\label{sec:Asymptotic-Analysis} Asymptotic Analysis }

Section \ref{subsec:algorithm-FS} has introduced forward selection
and then the $t$-statistic based on the selected variables. We proceed
by establishing asymptotic guarantee for this procedure. After laying
out the regularity conditions, we reverse the order by first studying
a generic post-selection inference in this context of ATE estimation,
and then arguing that the forward selection is a competitive method
for variable selection.

We work with a multi-index asymptotic framework. In asymptotic statements,
we take the number of cross-sectional units $N\to\infty$, while the
number of pre-treatment observations $T_{1}=T_{1}\left(N\right)$
is understood as a deterministic function of $N$ such that $T_{1}\to\infty$
as $N\to\infty$. $N$ is allowed to be larger than $T_{1}$ to accommodate
high-dimensional settings, although $\left(\log N\right)/T_{1}\to0$.
Similar indexing is applied to the number of the post-treatment observations
$T_{2}=T_{2}\left(N\right)\to\infty$ and $\left(\log N\right)/T_{2}\to0$
as $N\to\infty$. 

\subsection{Regularity Conditions for Pre-treatment Period}

The algorithm of forward selection uses the pre-treatment data only.
To study the asymptotic properties of forward selection and post-selection
inference, we impose two high-level assumptions. The first one regularizes
the minimal eigenvalue of the population Gram matrix. Let $\eta_{r}=\min_{\mathcal{U}_{r}}\phi_{\min}\left(\mathcal{E}_{\left(1\right)}\left[\mathbf{y}_{Ut}\mathbf{y}_{Ut}^{\prime}\right]\right)$
where $\mathcal{U}_{r}:=\left\{ U\subset\mathcal{N}:\left|U\right|\leq\left\lfloor r\right\rfloor \right\} $
for some $r\in\mathbb{\mathbb{R}}^{+}$. A \emph{universal constant}
is a strictly positive finite real number that is independent of sample
sizes.
\begin{assumption}
\label{assu:eigen} For any sequence $R=R\left(N\right)$ satisfying
$1/R+R/(T_{1}/\log N)^{1/3}\to0$ as $N\to\infty$, there are universal
constants $c$ and $\delta_{1}$ such that $\liminf_{N\to\infty}\eta_{(1+\delta_{1})R}\geq c$.
\end{assumption}
\begin{rem}
Stacking $\mathbf{y}_{\mathcal{N}_{0}t}^{0}:=(y_{jt}^{0})_{j\in\mathcal{N}_{0}}$,
we can write (\ref{eq:factor_single}) as an $\left(N+1\right)$-equation
system 
\begin{equation}
\mathbf{y}_{\mathcal{N}_{0}t}^{0}=\boldsymbol{\Lambda}\mathbf{f}_{t}+\mathbf{e}_{\mathcal{N}_{0}t},\ \ t\in\mathcal{T}\label{eq:factor_stack}
\end{equation}
where $\boldsymbol{\Lambda}:=(\lambda_{0},\lambda_{1},...,\lambda_{N})'$
is the $\left(N+1\right)\times K$ factor loading matrix and $\mathbf{e}_{\mathcal{N}_{0}t}:=(e_{jt})_{j\in\mathcal{N}_{0}}$
is the $\left(N+1\right)$-vector of zero mean idiosyncratic errors.
In the literature of large-dimensional factor models, \citet[p.141]{bai2003inferential}
assumes that $\phi_{\min}\left(\mathcal{E}_{\left(1\right)}\left[\mathbf{e}_{\mathcal{N}_{0}t}\mathbf{e}_{\mathcal{N}_{0}t}^{\prime}\right]\right)$
is bounded away from 0, which implies $\phi_{\min}\left(\mathcal{E}_{\left(1\right)}\left[\mathbf{y}_{\mathcal{N}_{0}t}\mathbf{y}_{\mathcal{N}_{0}t}^{\prime}\right]\right)$
is bounded away from 0 as well. Such a minimal eigenvalue condition
on the $\left(N+1\right)\times\left(N+1\right)$ population Gram matrix
is relaxed here in Assumption \ref{assu:eigen} to any $u\times u$
Gram submatrix with $u=\left|U\right|\leq(1+\delta_{1})R$. It echoes
the \emph{restricted eigenvalue condition }or the\emph{ compatibility
condition }that are routinely imposed in the high-dimensional regression
literature \citep[Section 6.13]{bickel2009simultaneous,buhlmann2011statistics}.
More precisely, our version is the \emph{sparse Riesz condition}\textbf{
}as in \citet{zhang2008sparsity} and \citet{chen2008extended}; while
these two papers set $\delta_{1}=1$, we allow any $\delta_{1}\in\left(0,1\right)$.
\end{rem}
As $R$ diverges to infinity at a rate slower than $\left(T_{1}/\log N\right)^{1/3}$,
the sample version of the $u\times u$ Gram submatrix $\mathbb{E}_{\left(1\right)}\left[\mathbf{y}_{Ut}\mathbf{y}_{Ut}^{\prime}\right]$
involving $T_{1}$ time series observations is likely to be of full
rank when $u\ll T_{1}$, with the help of the second assumption below
about the population second-moment as well as their sample counterpart.
\begin{assumption}
\label{assu:2nd-moment} 
\begin{enumerate}
\item $\max_{i,j\in\mathcal{N}_{0}}\left|\mathbb{E}_{\left(1\right)}[y_{it}y_{jt}]-\mathcal{E}_{\left(1\right)}[y_{it}y_{jt}]\right|=O_{p}(\sqrt{\left(\log N\right)/T_{1}}).$
\item $\max_{j\in\mathcal{N}_{0}}\mathcal{E}_{\left(1\right)}[y_{jt}^{2}]\leq C$
for a universal constant $C$.
\end{enumerate}
\end{assumption}
Assumption \ref{assu:2nd-moment}(a) postulates a convergence rate
of the second moments, and (b) is a common assumption of finite population
second moments. With independent observations, \citet{belloni2012sparse}
use the self-normalized Cram\'{e}r-type moderate-deviation theory
\citep{jing2003self} to establish the probabilistic bound in (a).
In time series contexts, similar conditions are used in \citet{medeiros2016l1},
\citet{kock2015oracle}, and \citet{koo2016high} under various assumptions
of tail bounds and serial dependence.

In the population model, $\boldsymbol{\beta}_{U}^{0}:=(\mathcal{E}_{\left(1\right)}[\mathbf{y}_{Ut}\mathbf{y}_{Ut}^{\prime}])^{-}\mathcal{E}_{\left(1\right)}\left[\mathbf{y}_{Ut}y_{0t}\right]$
is the ``true'' linear projection coefficient under a given $U$,
and the corresponding projection error is $\varepsilon_{Ut}:=y_{0t}-\mathbf{y}_{Ut}^{\prime}\boldsymbol{\beta}_{U}^{0}$.
Let $\sigma_{U}^{2}:=\mathcal{E}_{\left(1\right)}[\varepsilon_{Ut}^{2}]$
be the population variance of the projection error under the set $U$,
and $\widehat{\sigma}_{U}^{2}$ be the sample variance of $(\widehat{\varepsilon}_{Ut}:=y_{0t}-\mathbf{y}_{Ut}^{\prime}\widehat{\boldsymbol{\beta}}_{U})_{t\in\mathcal{T}_{1}}$.
The following lemma shows the sample variance $\hat{\sigma}_{U}^{2}$
approximates the population counterpart $\sigma_{U}^{2}$, and the
OLS estimator $\widehat{\boldsymbol{\beta}}_{U}=(\mathbb{E}_{\left(1\right)}[\mathbf{y}_{Ut}\mathbf{y}_{Ut}^{\prime}])^{-}\mathbb{E}_{\left(1\right)}\left[\mathbf{y}_{Ut}y_{0t}\right]$
approximates $\boldsymbol{\beta}_{U}^{0}$.
\begin{lem}
\label{lem:pop=000026sam} If Assumptions \ref{assu:eigen} and \ref{assu:2nd-moment}
hold, then
\begin{enumerate}
\item $\max\,_{\mathcal{U}_{(1+\delta_{1})R}}\ \left|\hat{\sigma}_{U}^{2}-\sigma_{U}^{2}\right|=O_{p}(\sqrt{R\left(\log N\right)/T_{1}})=o_{p}\left(1\right)$;
\item $\max\,_{\mathcal{U}_{(1+\delta_{1})R}}\ \Vert\widehat{\boldsymbol{\beta}}_{U}-\boldsymbol{\beta}_{U}^{0}\Vert_{2}=O_{p}(\sqrt{R^{3}\left(\log N\right)/T_{1}})=o_{p}\left(1\right).$
\end{enumerate}
\end{lem}
Lemma \ref{lem:pop=000026sam} indicates that over all index sets
$U$'s with no more than $\left\lfloor \left(1+\delta_{1}\right)R\right\rfloor $
elements, if $R$ diverges slowly such that $1/R+R/\left(T_{1}/\log N\right)^{1/3}\to0$
as in Assumption \ref{assu:eigen}, then the difference between $\widehat{\sigma}_{U}^{2}$
and $\sigma_{U}^{2}$ is negligible in probability. Similar approximation
holds in the coefficient estimation for $\boldsymbol{\beta}_{U}^{0}$.
These are results prepared for the following two subsections.

\subsection{\label{subsec:Generic-Post-Selection-Inference} Generic Post-Selection
Inference}

We first work on the asymptotic property of the post-selection $t$-statistic
based on a generic data-driven variable selection method using the
pre-treatment data only. It will include the forward selection as
a special case.

In inference we must use the post-treatment data, on which we impose
a few regularity assumptions.
\begin{assumption}
\label{assu:inference} 
\begin{enumerate}
\item $\max_{j\in\mathcal{N}_{0}}\left|\mathbb{E}_{\left(2\right)}[y_{it}^{0}]\right|=O_{p}\left(\sqrt{\left(\log N\right)/T_{2}}\right)$.
\item $\max_{i,j\in\mathcal{N}_{0}}\left|\mathbb{E}_{\left(2\right)}[y_{it}^{0}y_{jt}^{0}]-\mathcal{E}_{\left(2\right)}[y_{it}^{0}y_{jt}^{0}]\right|=O_{p}\left(\sqrt{\left(\log N\right)/T_{2}}\right).$
\item $\max_{t\in\mathcal{T}_{2},j\in\mathcal{N}_{0}}E[(y_{jt}^{0})^{4}]\leq C$.
\item $\liminf_{N\to\infty}\min_{U\subset\mathcal{N}}\ T_{2}^{-1}\sum_{t,s\in\mathcal{T}_{2}}E\left[\varepsilon_{Ut}\varepsilon_{Us}\right]\geq c$.
\item $\limsup_{N\to\infty}\max_{U\subset\mathcal{N}}\ T_{2}^{-1}\sum_{t,s\in\mathcal{T}_{2}}\left|E\left[\varepsilon_{Ut}\varepsilon_{Us}\right]\right|\leq C.$
\end{enumerate}
\end{assumption}
In the post-treatment subsample, Assumption \ref{assu:inference}(a)
is about the convergence rate of the sample mean to the population
mean 0, although $y_{0t}^{0}$ is unobservable. (b) is analogous to
Assumption \ref{assu:2nd-moment}(a) in the pre-treatment period,
and the fourth moment in (c) is commonly imposed in studies of inferential
procedures for high-dimensional factor models \citep{bai2003inferential}.
The last two items in Assumption \ref{assu:inference} are concerning
the long-run variance, where (d) bounds the long-run variance from
degeneracy and (e) guarantees the absolute summability of the autocorrelations.
(c), (d) and (e) make sure that the self-normalized test statistic
behaves well, so that a suitable version of the Berry-Essen bound
can be applied to establish the asymptotic normality of the test statistic.

Next, we introduce the time series weak dependence structure. Let
$\mathcal{F}_{N}^{t_{1},t_{2}}$ be the smallest $\sigma$-field generated
by the Borel sets of the collection $\{\left(\mathbf{f}_{t}',\mathbf{e}_{\mathcal{N}_{0}t}'\right)':t_{1}\leq t\leq t_{2}\}$
from the factor model (\ref{eq:factor_stack}), where it naturally
incorporates the fact that no random variables are produced at $t=0$,
the calendar date for the treatment. In view of the infinite time
series in Figure \ref{fig:timeline}, for each $k\in\mathbb{N}$ we
define
\begin{align}
\phi_{N}\left(k\right): & =\sup_{t\in\mathbb{Z}}\left\{ \left|\Pr\left(B|A\right)-\Pr\left(B\right)\right|:A\in\mathcal{F}_{N}^{-\infty,t},\ B\in\mathcal{F}_{N}^{t+k,\infty},\ \Pr\left(A\right)>0\right\} ,\label{eq:phi_coef}
\end{align}
 where $\mathbb{Z}$ is the set of all integers. The dependence indicator
$\phi_{N}\left(k\right)$ is the uniform strong mixing coefficient
\citep[p.209]{davidson1994stochastic}. We impose the following Assumption
\ref{assu:alpha_phi_coef}, which is similar to \citet{carvalho2016arco}'s
Assumption 3 of geometric strong mixing.
\begin{assumption}
\label{assu:alpha_phi_coef} There are two universal constants $c_{1}$
and $c_{2}$ such that $\limsup_{N\to\infty}\phi_{N}\left(k\right)\leq c_{1}\exp\left(-c_{2}k\right)$
for all $k\in\mathbb{N}$.
\end{assumption}
The above assumption is employed for two technical purposes: (i) It
allows us to invoke the Berry-Essen bound for heterogeneous time series
\citep{bentkus1997berry,sunklodas2000approximation}. (ii) It implies
the asymptotic independence, as $k\to\infty$, between the events
in $\mathcal{F}_{N}^{-\infty,-1}$ before the treatment and the events
in $\mathcal{F}_{N}^{k,\infty}$ which is $k$ periods after the treatment.
The second point is critical for asymptotic normality. If a single
dataset is used for model selection and parameter estimation, post-selection
inference is in general a very difficult statistical problem that
leads to non-standard asymptotic distributions \citep{leeb2005model,leeb2006can},
and it is a topic of intensive recent research \citep{berk2013valid,belloni2014uniform,belloni2017program,hansen2018targeted}.
However, in conditional (on the selected model from a training sample)
predictive inference, post-selection asymptotic normality is achievable
\citep{leeb2009conditional} and the inference can be carried out
following standard asymptotically normal procedure. 

In our context, the estimated ATE is based on the average involving
the predicted outcomes over the post-treatment period $\mathcal{T}_{2}$.
Between the two blocks $\mathcal{T}_{1}$ and $\mathcal{T}_{2}$,
the observations near the treatment date $t=0$ are essentially dependent.
For instance, those with time index $t=1,\ldots,k$ are statistically
dependent on the random variables at the end of $\mathcal{T}_{1}$.
This dependent episode consists of a smaller and smaller fraction
of the post-treatment sample if we devise $k=k\left(N\right)$ such
that $k/T_{2}\to\infty$ as $N\to\infty$.

Let $M$ be the DGP that generates $\mathbf{y}_{\mathcal{N}_{0}t}^{0}$
in (\ref{eq:factor_stack}). Let $\check{U}_{R}\in\mathcal{U}_{R}$
be an index set estimated by an arbitrary variable selection method
using the pre-treatment dataset only. Let $\mathcal{Z}_{\check{U}_{R}}:=\mathcal{Z}_{U}|_{U=\check{U}_{R}}$
for the $t$-statistic $\mathcal{Z}_{U}$ defined in (\ref{eq:z-test})
evaluated at $U=\check{U}_{R}$. Obviously $\mathcal{Z}_{\check{U}_{R}}=\mathcal{Z}_{\check{U}_{R}}\left(M\right)$
depends on the underlying DGP $M$.

Consider a set of DGPs $\mathcal{M}$ such that Assumptions \ref{assu:eigen},
\ref{assu:2nd-moment}, \ref{assu:inference} and \ref{assu:alpha_phi_coef}
hold \emph{uniformly}. This uniformity requirement strengthens the
stochastic orders in Assumptions \ref{assu:2nd-moment}(a), \ref{assu:inference}(a)
and (b), whereas all other assumptions are already stated with universal
constants. The following theorem provides the asymptotic distribution
of $\mathcal{Z}_{\check{U}_{R}}$ uniformly over the set of eligible
DGPs $M\in\mathcal{M}$.
\begin{thm}
\label{thm:uniform_inference} If $T_{1}^{-1}R^{4}\log^{2}N\log^{4}T_{2}\to0$
and $1/\tau+\tau/\log T_{2}\to0$ as $N\to\infty$, then under the
null hypothesis $\mathbb{H}_{0}$ we have 
\[
\sup_{M\in\mathcal{M}}\left|\Pr\big(\mathcal{Z}_{\check{U}_{R}}\left(M\right)\leq a\big)-\Phi\left(a\right)\right|\to0\ \ \text{for all }a\in\mathbb{R}.
\]
\end{thm}
The restriction $M\in\mathcal{M}$ implicitly imposes Assumptions
\ref{assu:eigen}, \ref{assu:2nd-moment}, \ref{assu:inference} and
\ref{assu:alpha_phi_coef} uniformly. Theorem \ref{thm:uniform_inference}
is established by a Berry-Essen bound for heterogeneous time series
\citep{sunklodas2000approximation}. The key condition that contributes
to the uniform asymptotic normality is the weak dependence in Assumption
\ref{assu:alpha_phi_coef} along with our setting of ATE estimation,
in which the policy intervention occurs at time $t=0$ splits the
sample into two disjoint subsamples indexed by $\mathcal{T}_{1}$
and $\mathcal{T}_{2}$, respectively. Sampling splitting is a popular
approach to achieve uniform inference in statistical machine learning
in cross-sectional environments, for example \citet{belloni2014uniform}
and \citet{wager2018estimation}. The notation of uniform strong mixing
is a time series analogy of asymptotic independence.

\begin{example}\label{exa:theory}
Theorem \ref{thm:uniform_inference} holds no matter whether the coefficient
$\boldsymbol{\beta}_{\mathcal{N}}^{0}$ in (\ref{eq:y0}) is sparse
or not. Consider a regression equation $y_{0t}=\sum_{j\in\mathcal{N}}\beta_{j}^{0}y_{jt}+\varepsilon_{t}$
where the regressor $y_{jt}\sim\mathrm{iid}\ N\left(0,1\right)$ across
$j\in\mathcal{N}$ and $t\in\mathcal{T}_{1}$, the coefficient $\beta_{j}^{0}=1/\sqrt{N}$,
and the error term $\varepsilon_{t}\sim\mbox{iid}\ N\left(0,\sigma_{\varepsilon}^{2}\right)$
is independent of the regressors. Since $\beta_{j}^{0}$ is of order
$N^{-1/2}$ for all $j$ here, this is an extremely dense regression
model; when $N/T_{1}\to\infty$, it is impossible to accurately estimate
all these coefficients. Theorem \ref{thm:uniform_inference} is immune
from the dense model estimation difficulty because it is sufficient
if we can approximate the lower-dimensional vector $\boldsymbol{\beta}_{U}^{0}|_{U=\check{U}_{R}}$
well enough, instead of the intractable high-dimensional $N$-vector
$\boldsymbol{\beta}_{\mathcal{N}}^{0}$. This example will be continued
after presenting Theorem \ref{thm:FS_var_min} later.
\end{example}

The uniform asymptotic normality in Theorem \ref{thm:uniform_inference}
holds regardless of the algorithm that selects a subset of no more
than $R$ control variables. Consider an \emph{ad hoc} non-random
way of choosing a sequence of sets. Given an arbitrary ordering of
the control units, we may naively choose the first $R$ terms $U_{R}^{\mathrm{naive}}=\left\{ 1,\ldots,R\right\} $
for $R$ satisfying the order regularized by the conditions in Assumption
\ref{thm:uniform_inference}, and we would have $\mathcal{Z}_{U_{R}^{\mathrm{naive}}}\stackrel{d}{\to}N\left(0,1\right)$.
It is also applicable to the $t$-statistic based on HCW's best subset
method via AIC or AICC. When they developed the asymptotic inference,
HCW heuristically took the selected variables, which we denote here
as $\check{U}_{R}^{\mathrm{AICC}}$, as if they were fixed. Our result
implies $\mathcal{Z}_{\check{U}_{R}^{\mathrm{AICC}}}\stackrel{d}{\to}N\left(0,1\right)$,
which helps justify HCW's practice. 

Instead of $\check{U}_{R}^{\mathrm{AICC}}$ that is based on exhaustive
search over all subsets, we nevertheless advocate the forward selection
algorithm for $\widehat{U}_{R}$ in view of its convenience in computation
in high-dimensional settings. The asymptotic theory of forward selection
is developed in the next subsection.

\subsection{\label{subsec:eff-of-Forward} Efficacy of Forward Selection}

In order to discuss the efficacy of forward selection, we spell out
our target for variable selection. Let $U_{u}^{*}:=\arg\min_{\mathcal{U}_{u}}\sigma_{U}^{2}$
be the \emph{best subset} of $u$ elements among all $U\subset\mathcal{N}$;
the number of elements is no more than some $u\in\mathbb{N}$. Let
$\sigma_{u}^{*2}:=\sigma_{U_{u}^{*}}^{2}=\sigma_{U}^{2}|_{U=U_{u}^{*}}$
be the corresponding noise level under this best subset $U_{u}^{*}$.
If $U_{u}^{*}$ is not unique, we simply refer to any of them as the
best subset and our analysis is not affected no matter $U_{u}^{*}$
is unique or not. It is computationally expensive to locate the best
subset $U_{u}^{*}$. Even if $\sigma_{U}^{*2}$ were estimated with
no noise, we would exhaustively compare $\sigma_{U}^{*2}$ for ${N \choose u}$
models, which is of \emph{exponential} order of $N$.

Instead of searching for $U_{u}^{*}$, we seek to identify a subset
$U$ on which $\sigma_{U}^{2}$ approximates the optimal variance
$\sigma_{u}^{*2}$. Theorem \ref{thm:FS_var_min} below states that
the greedy forward selection algorithm picks up a set $\widehat{U}_{R}$
with a regression variance asymptotically as small as the desired
$u$-element best subset if $R$ dominates $u$ asymptotically. The
greedy algorithm only searches among $\sum_{r=1}^{R}\left(N-r+1\right)$
models, which is of \emph{linear} order of $N$. The latter is computationally
much more efficient than exhaustive search. 
\begin{thm}
\label{thm:FS_var_min} Suppose Assumptions \ref{assu:eigen} and
\ref{assu:2nd-moment} hold. For any sequence $u=u\left(N\right)$
such that $u/R\to0$ as $N\to\infty$, we have 
\[
\Pr\left(\widehat{\sigma}_{\widehat{U}_{R}}^{2}\leq\sigma_{u}^{*2}+\delta_{2}\right)\to1
\]
 for any universal constant $\delta_{2}>0$.
\end{thm}
Since variable selection does not use the post-treatment subsample,
only Assumptions \ref{assu:eigen} and \ref{assu:2nd-moment} are
needed for Theorem \ref{thm:FS_var_min}. The above theorem is a nearly
optimal result. It implies with high probability that the computationally
feasible sample variance $\widehat{\sigma}_{\widehat{U}_{R}}^{2}$
is asymptotically no worse, up to an arbitrarily small tolerance $\delta_{2}$,
than the computationally heavy but theoretically optimal $\sigma_{u}^{*2}$.
Such approximation can be achieved by incorporating $R$ units. Though
$R$ is of bigger order than $u$ in the asymptotic sense, if we specify
$R=\left\lfloor u\log\log N\right\rfloor $, then obviously the number
of OLS regressions is fewer than $Nu\log\log N$, and $Nu\log\log N\ll{N \choose u}$
for a non-trivial $u$ and large $N$.

\begin{example*}
(Example \ref{exa:theory} continues.) For the dense model in our example, when $R\ll N$
there must be non-trivial gap between $\min_{U\in\mathcal{U}_{R}}\{\sigma_{U}^{2}\}$
and $\sigma_{\varepsilon}^{2}=\sigma_{\mathcal{N}}^{2}$, where the
latter can be achieved only when all the control variables are selected
and is infeasible in the high-dimensional setting. Nevertheless, according
to Theorem \ref{thm:FS_var_min}, the forward selection algorithm
will pick an $R$-regressor model that dominates the optimal set $U_{u}^{*}$
in terms of the associated population variances even if $u\to\infty$
as $N\to\infty$, provided $R/u\to\infty$.

\end{example*}
\begin{rem}
If the best subset $U_{u}^{*}$ is sparse, for example in a sparse
linear regression with only a few non-zero coefficients, Theorem \ref{thm:FS_var_min}
may not be surprising as these non-zero coefficients will all be selected
with high probability. The novelty of this result lies in that it
imposes no sparsity assumption on the regression coefficients. The
result relies on Assumption \ref{assu:eigen}, which is a natural
implication of standard factor models in high dimensional \citep{bai2003inferential}.
One of the key steps in the proof of Theorem \ref{thm:FS_var_min}
is Lemma \ref{lem:inequality} in the Appendix, based on the submodularity
ratio studied by \citet{das2011submodular} for greedy algorithms
in the population model. To accommodate sampling errors, in Lemma
\ref{lem:population} in the Appendix we introduce a sequence of sets
with a tolerance. The theoretical results that link the sample to
the population go beyond the coverage of \citet{das2011submodular}.
\end{rem}
Theorem \ref{thm:FS_var_min} holds uniformly if the DGPs under consideration
are restricted to $\mathcal{M}$. After forward selection, we use
$\mathbf{y}_{\widehat{U}_{R}t}$ to predict the counterfactual $y_{0t}^{0}$
and obtain the time-varying treatment effect $\hat{\Delta}_{\widehat{U}_{R}t}$
in the post-treatment period. Since $\mathcal{Z}_{\hat{U}_{R}}\left(M\right)$
is a special case of $\mathcal{Z}_{\check{U}_{R}}\left(M\right)$
in Theorem \ref{thm:uniform_inference} when we use forward selection
to choose variables, the following corollary is an immediate implication.
\begin{cor}
\label{cor:FS}Under the conditions in Theorem \ref{thm:uniform_inference},
the $t$-statistic based on the estimated set $\widehat{U}_{R}$ by
forward selection satisfies
\[
\sup_{M\in\mathcal{M}}\left|\Pr\left(\mathcal{Z}_{\widehat{U}_{R}}\left(M\right)\leq a\right)-\Phi\left(a\right)\right|\to0\ \ \text{for all }a\in\mathbb{R}.
\]
\end{cor}
We summarize the theoretical results in this paper. Theorem \ref{thm:uniform_inference}
shows that the $t$-statistic based on a generic variable selection
method from the pre-treatment data has correct size. Theorem \ref{thm:FS_var_min}
highlights that the forward selection algorithm can attain variance
for the regression model as small as that of the best subset $U_{u}^{*}$
if $R$ dominates the cardinality $u$ asymptotically. The small $\widehat{\sigma}_{\widehat{U}_{R}}^{2}$
in general improves the statistical efficiency of the hypothesis testing.
\begin{rem}
\label{rem:lasso} Before we go to the simulation exercises, we comment
on the distinctions between forward selection and Lasso. Forward selection
explicitly controls the number of variables included in the regression
and the regression coefficients are estimated by OLS. On the other
hand, Lasso estimates the parameter by
\[
\hat{\boldsymbol{\beta}}_{\lambda}^{las}:=\arg\min_{\boldsymbol{\beta}\in\mathbb{R}^{N}}\left\{ \mathbb{E}_{\left(1\right)}\left[\left(y_{0t}-\mathbf{y}_{\mathcal{N}t}^{\prime}\boldsymbol{\beta}\right)^{2}\right]+\lambda\left\Vert \boldsymbol{\beta}\right\Vert _{1}\right\} ,
\]
where $\lambda$ is the penalty level tuning parameter. Usually the
asymptotic theory for Lasso is derived when $\lambda$ slowly shrinks
to 0 at some rate as $N\to\infty$, which does not explicitly control
the number of selected variables. Moreover, standard Lasso theory
assumes sparsity in the $N$-vector $\boldsymbol{\beta}_{\mathcal{N}}^{0}$,
as in \citet{carvalho2016arco}'s Assumption 4, and gives the rate
of convergence of some vector norm of $\hat{\boldsymbol{\beta}}_{\lambda}-\boldsymbol{\beta}_{\mathcal{N}}^{0}$.
\citet[p.410]{efron2004least} explore the algorithmic connections
between them, and demonstrate that Lasso is a less aggressive selection
strategy than forward selection. This remark will be continued in
the next section after we report the simulation results.
\end{rem}

\section{Simulations\label{sec:Simulations}}

We evaluate the finite-sample performance of our proposed procedure
by Monte Carlo simulations. We conduct extensive experiments with
non-sparse coefficients and sparse ones, and with various degrees
of cross-sectional correlation and time dependence.\footnote{Due to the limitations of space, in the main text we present results
for a dense underlying linear regression model. In the online supplement,
we document the performance of variable selection, parameter estimation
and prediction accuracy for a sparse model.} For comparison, we also estimate the model using Lasso. For each
DGP, we generate one treated unit $j=0$ along with 100 control units
$j=1,\ldots,100$. We run $1000$ replications and check the out-of-sample
root mean predicted squared error (RMPSE) as well as the test size
or power for $\mathbb{H}_{0}$. For simplicity, we set equal the lengths
of the pre-treatment and post-treatment time series, with $T_{1}=T_{2}=50,100$
or 200. 

Both forward selection and Lasso need turning parameters: the stopping
time $R$ in forward selection and the penalty level $\lambda$ in
Lasso. We adopt the modified BIC \citep{wang2009shrinkage} in choosing
the tuning parameters. For forward selection, the stopping time $R$
is determined by 
\[
\widehat{R}=\arg\min_{r\in\mathbb{N}}\left\{ \log\left(\hat{\sigma}_{\widehat{U}_{r}}^{2}\right)+\log\log N\cdot r(\log T_{1})/T_{1}\right\} .
\]
 Lasso's tuning parameter is determined by
\[
\widehat{\lambda}=\arg\min_{\lambda}\left\{ \log\left(\mathbb{E}_{\left(1\right)}\left[(y_{0t}-\mathbf{y}_{\mathcal{N}t}'\hat{\boldsymbol{\beta}}_{\lambda}^{las})^{2}\right]\right)+2\log\log N\cdot\Vert\hat{\boldsymbol{\beta}}_{\lambda}^{las}\Vert_{0}(\log T_{1})/T_{1}\right\} ,
\]
where $\Vert\hat{\boldsymbol{\beta}}_{\lambda}^{las}\Vert_{0}$ is
the number of non-zero coordinates in the vector $\hat{\boldsymbol{\beta}}_{\lambda}^{las}$.
In the second term of the modified BIC, we have the admittedly \emph{ad
hoc} constant 1 for forward selection and 2 for the Lasso, respectively.
The difference arises because in our simulations Lasso would select
many more variables than forward selection were the same constant
shared in the two estimation methods, resulting in unsatisfactory
performance. The choice of the constant will be commented in the continued
Remark \ref{rem:lasso} in this section.

\subsection{Data Generating Processes}

We generate the data via the factor model (\ref{eq:factor_stack})
with 4 common factors. The idiosyncratic shocks $e_{jt}\sim N\left(0,0.5^{2}\right)$
is independent across $j$ and $t$.
\begin{itemize}
\item (i.i.d.~factors) All factors $f_{lt}\sim\mathrm{i.i.d.}\ N\left(0,l^{2}\right)$
across $t\in\mathcal{T}$ and $l=1,\ldots,4$. This DGP serves as
a benchmark.
\item (time-dependent factor) The dynamic factors are
\begin{align*}
\mathrm{iid:}\ \ f_{1t} & =u_{1t}\\
\mathrm{AR(1):}\ \ f_{2t} & =0.9f_{2,t-2}+u_{2t}\\
\mathrm{MA(2):}\ \ f_{3t} & =u_{3t}+0.8u_{3t-1}+0.4u_{3t-2}\\
\mathrm{ARMA(1,1):}\ \ f_{4t} & =0.5f_{4,t-1}+u_{4t}+0.5u_{4t-1}
\end{align*}
for $t\in\mathcal{T}$ where $u_{lt}\sim N\left(0,1\right)$ independently
across $t$ and $l$.
\end{itemize}
The factor loading $\lambda_{jl}$, $l=1,\ldots,4$, is independently
drawn from $\mathrm{Uniform}\left(1,2\right)$ if $j=0,\ldots,4$,
whereas $\lambda_{jl}\sim\mathrm{Uniform}\left(-0.1,0.1\right)$ if
$j=5,\ldots,100$. 

For $t\in\mathcal{T}_{2}$, the treated unit $y_{0t}$ is subject
to an exogenous shock $\Delta_{t}$. We generate $\Delta_{t}$ by
seven DGPs, denoted as $D1$ to $D7$:
\begin{gather*}
D1:\Delta_{t}=0;\ \ D2:\Delta_{t}\sim N(0,1);\ \ D3:\Delta_{t}=0.3\Delta_{t-1}+w_{t},\ w_{t}\sim N(0,1)\\
D4:\Delta_{t}\sim N(0.5,1);\ \ D5:\Delta_{t}\sim N(1,1)\\
D6:\Delta_{t}=0.35+0.3\Delta_{t-1}+w_{t},\ w_{t}\sim N(0,1);\ \ D7:\Delta_{t}=0.7+0.3\Delta_{t-1}+w_{t},\ w_{t}\sim N(0,1).
\end{gather*}
The null hypothesis is true under $D1$\textendash $D3$, but false
under $D4$\textendash $D7$. The treatment is time-invariant under
$D1$, time-varying under $D2$, and serially correlated under $D3$.
Mean shifts are introduced to post-treatment outcomes in $D4$ and
$D5$, whereas $D6$ and $D7$ add non-zero dynamic treatment effects.

\subsection{Implementation and Results}

The first two columns of Table \ref{tab:MSE_ns} report the number
of non-zero coefficients and the empirical RMPSE $\big(\mathbb{E}_{\left(2\right)}[\left(y_{0t}^{0}-\hat{y}_{0t}\right)^{2}]\big)^{1/2}$,
where $\hat{y}_{0t}$ is the predicted value for $y_{0t}$: forward
selection gives $\hat{y}_{0t}=\hat{y}_{0t}(\widehat{U}_{R})=\mathbf{y}_{\widehat{U}_{\widehat{R}}t}^{\prime}\hat{\boldsymbol{\beta}}_{\widehat{U}_{\widehat{R}}}$
and Lasso gives $\hat{y}_{0t}=\hat{y}_{0t}(\hat{\boldsymbol{\beta}}_{\widehat{\lambda}}^{las})=\mathbf{y}_{\mathcal{N}t}'\hat{\boldsymbol{\beta}}_{\widehat{\lambda}}^{las}$.
In both factor structures RMPSE of Lasso are larger than those of
forward selection in all cases, and Lasso chooses more variables.

\begin{table}[th]
\caption{\label{tab:MSE_ns}Variable Selection, RMPSE and Rejection Probabilities}
\medskip{}

\begin{centering}
\begin{tabular}{lrcc|ccc|cccc}
\hline 
 &  & \multicolumn{1}{r}{No.~of Sel.} & \multicolumn{1}{c|}{RMPSE} & \multicolumn{3}{c|}{Test Size} & \multicolumn{4}{c}{Test Power}\tabularnewline
 & $\left(T_{1},T_{2}\right)$ & variables &  & $D1$ & $D2$ & $D3$ & $D4$ & $D5$ & $D6$ & $D7$\tabularnewline
\hline 
\multicolumn{11}{c}{Forward selection}\tabularnewline
\multirow{3}{*}{\begin{turn}{90}
i.i.d.
\end{turn}} & $50$ & 6 & 0.813 & 0.066 & 0.066 & 0.099 & 0.785 & 1.000 & 0.664 & 0.992\tabularnewline
 & $100$ & 7 & 0.710 & 0.059 & 0.057 & 0.084 & 0.983 & 1.000 & 0.908 & 1.000\tabularnewline
 & $200$ & 9 & 0.656 & 0.059 & 0.055 & 0.077 & 1.000 & 1.000 & 0.995 & 1.000\tabularnewline
\hline 
\multirow{3}{*}{\begin{turn}{90}
dyn.
\end{turn}} & $50$ & 6 & 0.815 & 0.115 & 0.087 & 0.112 & 0.756 & 0.998 & 0.636 & 0.986\tabularnewline
 & $100$ & 7 & 0.710 & 0.088 & 0.070 & 0.091 & 0.975 & 1.000 & 0.892 & 1.000\tabularnewline
 & $200$ & 8 & 0.657 & 0.069 & 0.059 & 0.079 & 1.000 & 1.000 & 0.994 & 1.000\tabularnewline
\hline 
% &  &  & \multicolumn{1}{c}{} &  &  & \multicolumn{1}{c}{} &  &  &  & \tabularnewline
\multicolumn{11}{c}{Lasso}\tabularnewline
\multirow{3}{*}{\begin{turn}{90}
iid
\end{turn}} & $50$ & 9 & 0.968 & 0.063 & 0.067 & 0.096 & 0.724 & 0.998 & 0.622 & 0.985\tabularnewline
 & $100$ & 11 & 0.842 & 0.058 & 0.057 & 0.081 & 0.964 & 1.000 & 0.881 & 1.000\tabularnewline
 & $200$ & 14 & 0.739 & 0.056 & 0.055 & 0.078 & 1.000 & 1.000 & 0.993 & 1.000\tabularnewline
\hline 
\multirow{3}{*}{\begin{turn}{90}
dyn.
\end{turn}} & $50$ & 6 & 1.046 & 0.244 & 0.191 & 0.180 & 0.618 & 0.940 & 0.513 & 0.899\tabularnewline
 & $100$ & 8 & 0.902 & 0.184 & 0.146 & 0.126 & 0.870 & 0.998 & 0.775 & 0.994\tabularnewline
 & $200$ & 13 & 0.746 & 0.116 & 0.095 & 0.089 & 0.996 & 1.000 & 0.978 & 1.000\tabularnewline
\hline 
\end{tabular}
\par\end{centering}
\vspace{0.2cm}\footnotesize Notes: ``i.i.d.''~is short for the
i.i.d.~factor structure and ``dyn.''~for the dynamic factor structure.
The sample size is the number of $T_1$, and $T_2 = T_1$.
The first column ``No.~of Sel.~varaibles'' is the median of the
number of selected control units over the replications, and ``RMPSE''
is the empirical RMPSE over the replications. The entries for $D1$-$D3$
display the test size and those for $D4$-$D7$ show the power. The
nominal size test size is 5\%, and the empirical rejection probability
is computed over the replications.
\end{table}

\begin{rem*}
(Remark \ref{rem:lasso} continues.) Forward selection and Lasso differ
in their ways of coefficient estimation. If they are given the same
set of active variables, the resulting $\widehat{\sigma}^{2}$ from
forward selection is smaller than that of Lasso, because forward selection
estimates the coefficients by OLS whereas Lasso squeezes the coefficients
toward zero via the $L_{1}$ shrinkage. When estimation does not overfit
in the pre-treatment data, the aggressiveness of forward selection
contributes to the smaller RMPSE for the counterfactual in the post-treatment
subsample. Had Lasso selected the same number of variables as forward
selection, Lasso's RMPSE would be even worse than those in Table \ref{tab:MSE_ns}.
Thus in our simulations we tune the constants in the modified BIC
to allow Lasso to take in more variables in order to compensate Lasso's
restricted coefficient estimation.
\end{rem*}
\begin{rem}
In general, if the goal of variable selection is to identify a few
important and potentially causal variables to interpret the outcome,
we recommend Lasso or, even better, the adaptive Lasso \citep{zou2006adaptive}
which enjoys variable selection consistency. On the other hand, forward
selection is a competitive method in high-dimensional problems if
the purpose is synthesizing an ensemble of variables to mimic the
outcome but the identities of the selected variables are not of interest.
PDA matches the second purpose well.
\end{rem}
Columns 3\textendash 9 of Table \ref{tab:MSE_ns} display the rejection
probability of the null hypothesis, that is, the proportion of instances
when the test rejects the null. The nominal test size is 5\%. As the
null hypothesis is true in $D1$\textendash $D3$, the rejection probability
is associated with test size; the closer it approaches to $5\%$,
the better is the performance. For $D4$\textendash $D7$, on the
contrary, the larger is the rejection probability, the more powerful
is the test. We observe that as the length of the time series increases,
the test size based on forward selection falls down toward $5\%$
under both the static and dynamic factor structures, though there
is a slight size inflation in $D3$ when dynamics is present in the
factors. This is caused by the relatively imprecise long-run variance
estimation. The test is powerful under $D4$\textendash $D7$ when
the null is violated. In contrast, the test size of the model selected
by Lasso is subject to more severe size inflation and is less powerful.
The inferior test performance is caused by Lasso's larger RMPSE, which
is further caused by the shrinkage estimation scheme.\footnote{Simulation evidence of Lasso's coefficient estimation bias is shown
in Table S4 in the online supplement for a sparse model.}

\begin{figure}
\subfloat[i.i.d~factors]{
\centering
\includegraphics[scale=0.5]{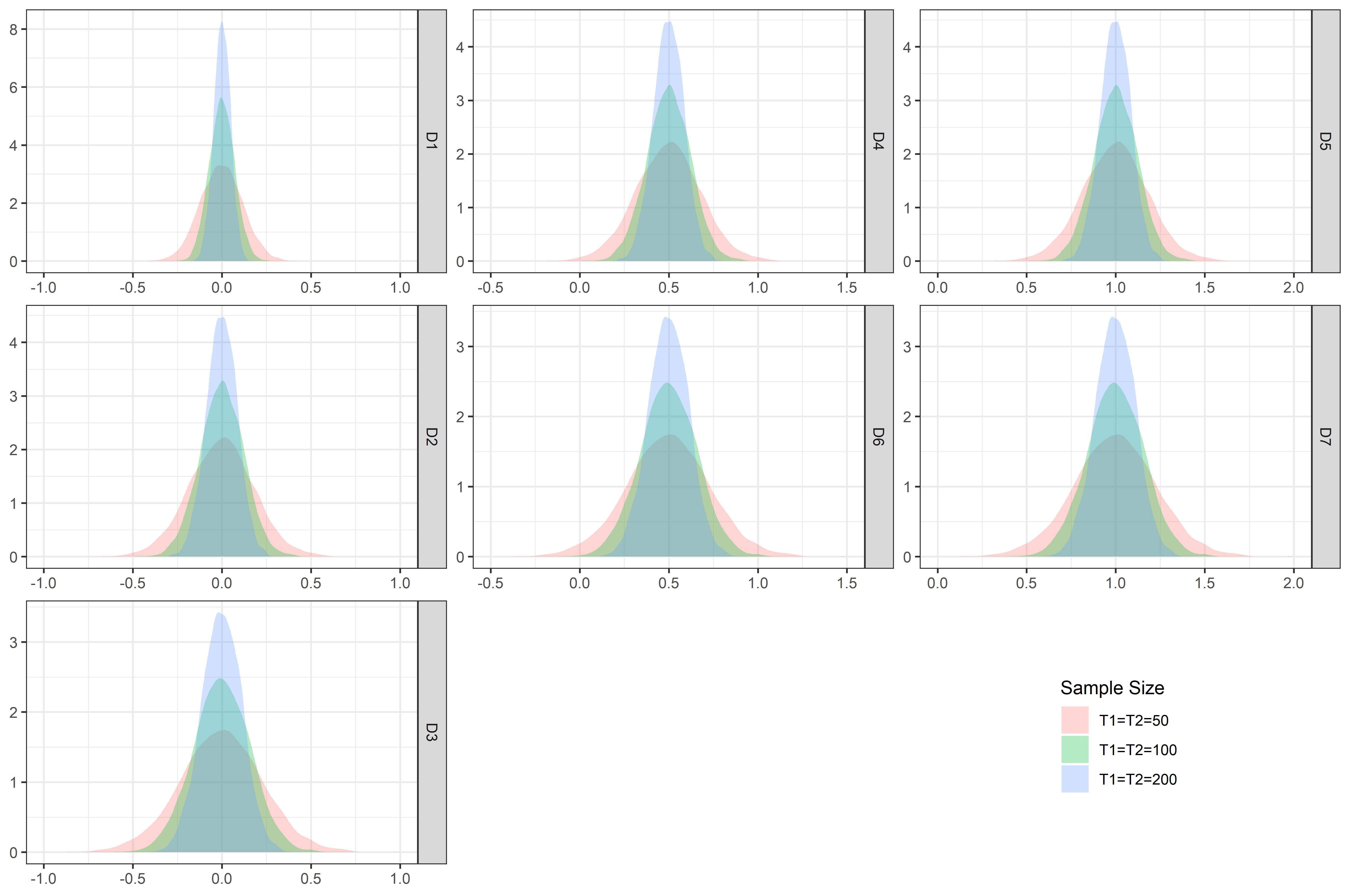}}

\subfloat[dynamic factors]{
\centering
\includegraphics[scale=0.5]{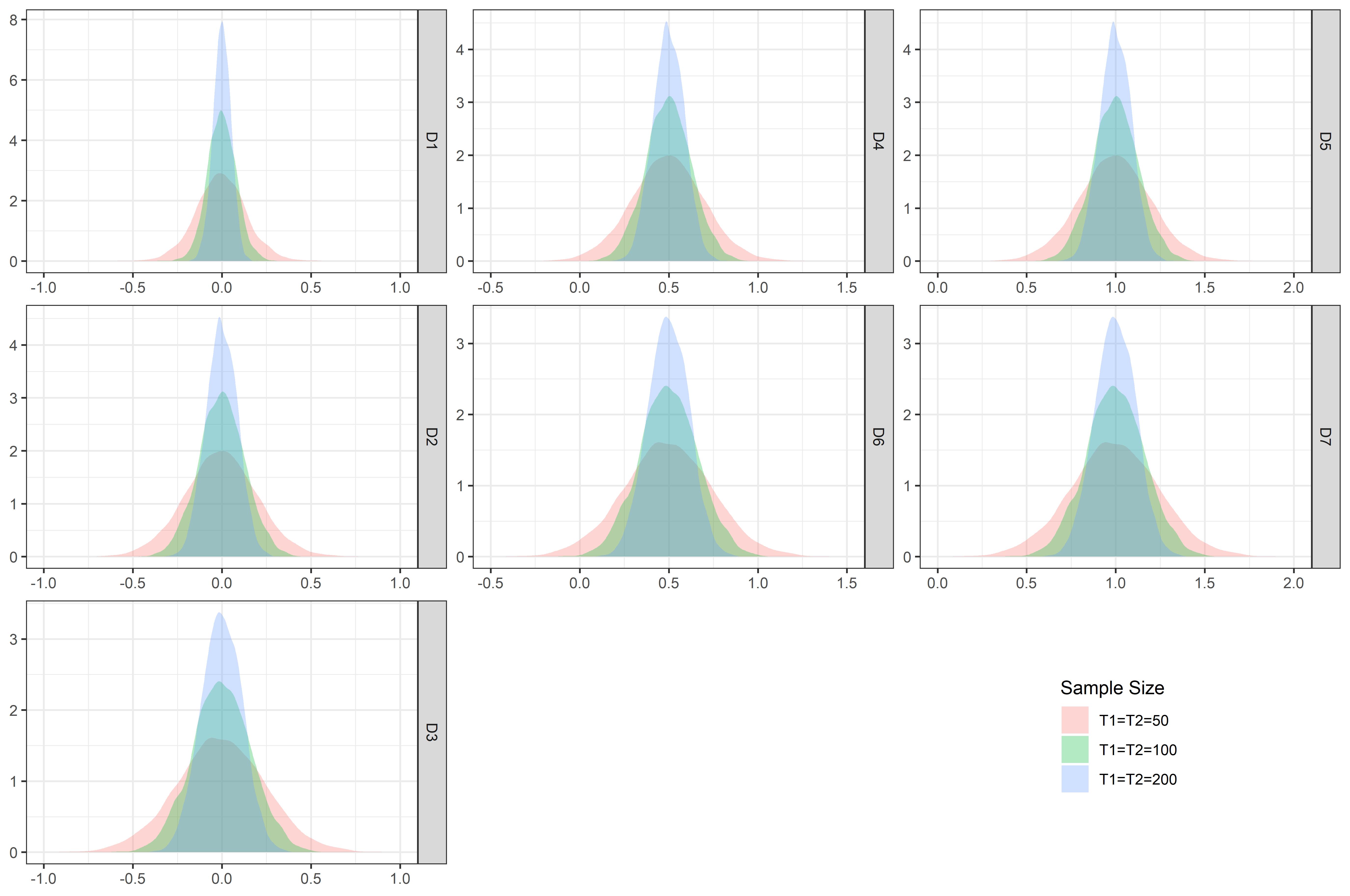}}

\caption{\label{fig:ate-bst}Kernel Density of the Estimated ATE }
\end{figure}

\begin{figure}[th]
\begin{centering}
\includegraphics[scale=0.5]{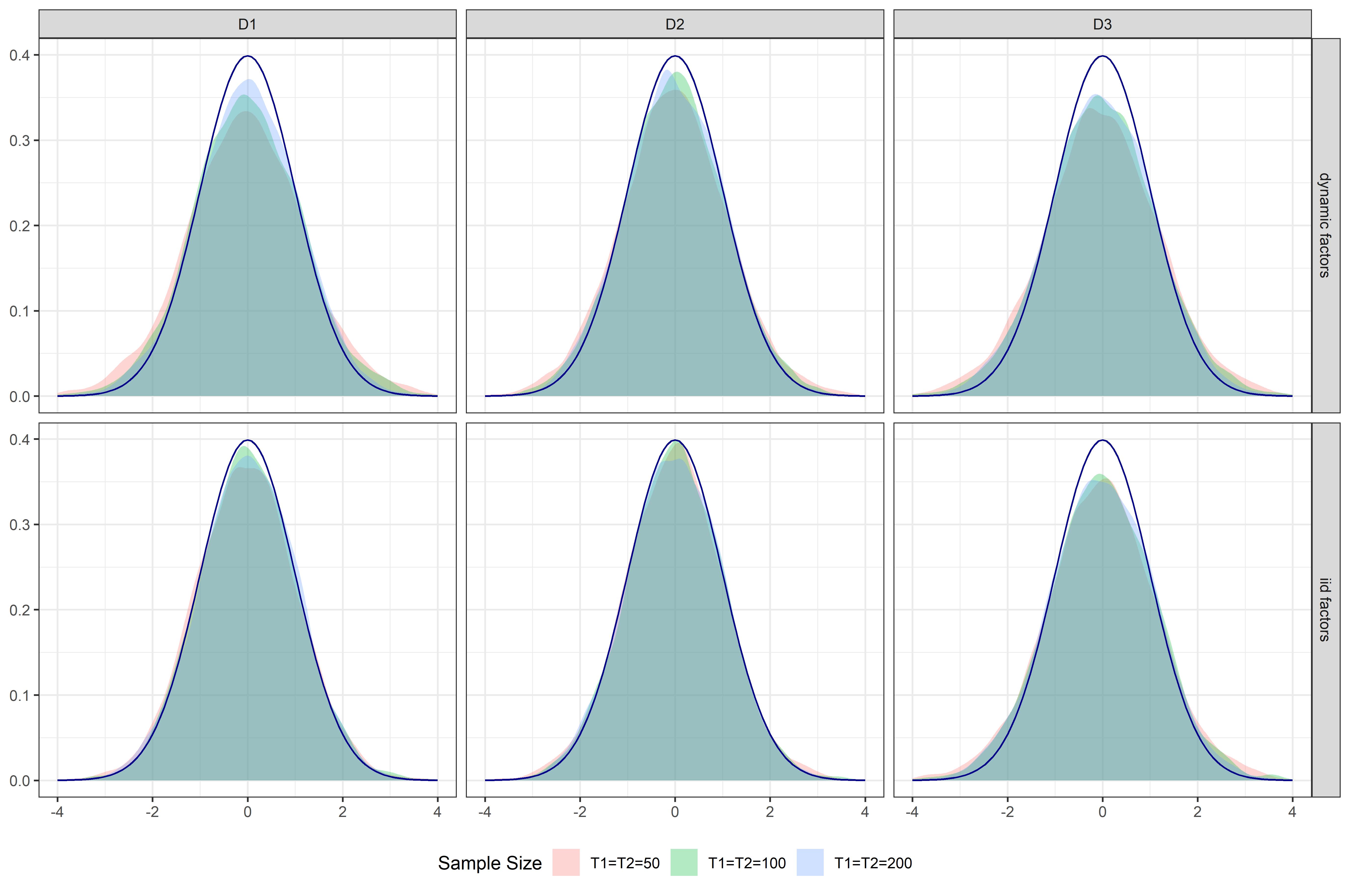}
\par\end{centering}
\begin{raggedright}
\vspace{0.2cm}\footnotesize Notes: The blue bell-shape curve is the
density of the standard normal distribution $N\left(0,1\right)$,
which is the limiting distribution of the $t$-statistic.
\par\end{raggedright}
\centering{}\caption{\label{fig:pdf-size}Kernel Density of the Test Statistic Under the
Null}
\end{figure}

We plot in Figure \ref{fig:ate-bst} the estimated ATE to facilitate
visualization. In each panel, the null hypothesis is true for the
first column of subgraphs, whereas the null is violated with $E\left[\Delta_{t}\right]=0.5$
for all $t\in\mathcal{T}_{2}$ in the second column and $E\left[\Delta_{t}\right]=1$
in the last column. We witness in both factor structures that forward
selection estimates the counterfactual with little bias and the variance
is reduced as the time length grows. Finally, the kernel density of
the test statistic $\mathcal{Z}_{\widehat{U}_{\widehat{R}}}$ based
on forward selection is shown in Figure \ref{fig:pdf-size}. Normality
is approximated very well in $D1$ and $D2$, though slightly heavier
tails are observed in $D3$. Overall, the $t$-statistic graph is
supportive for the theoretical result of asymptotic normality.

\section{Empirical Application\label{sec:Empirical-Applications}}

In this section, we investigate an empirical example where the number
of potential control units overpasses the number of temporal observations.
Another application which revisits HCW's original empirical example
is included in Section S2 of the online supplement. 

\subsection{Background and Data}

China launched an anti-corruption campaign of unprecedented scale
in November 2012 shortly after Xi Jinping took office. The campaign
aimed at cracking down graft and power abuse in all party apparatus,
government bureaucracies and military departments. The influence of
the anti-corruption campaign motivates academic research assessing
its impact from a multitude of perspectives, for example, stock return
\citep{lin2016anti,ding2017equilibrium} and corporate behavior \citep{xu2016does,PAN2017}.
In this paper, we investigate luxury goods importation.

We use the import data from the United Nations Comtrade Database.\footnote{The United Nations Statistics Division, United Nations Comtrade database.
\url{http://comtrade.un.org/}.} The database provides detailed statistics for international commodity
trade, and the monthly data for China are available since 2010. We
focus on the category named ``watches with case of, or clad with,
precious metal,'' following \citet{lan2018swiss} who find that Chinese
luxury watches import co-moves with leadership transitions and government
turnover. 

The raw time series of Chinese luxury watch import, plotted as the
red curve in the lower subgraph in Figure \ref{fig:Luxury-Watches-Import:},
dropped sharply around the start of the anti-corruption campaign.
However, a seemingly structural break can be the upshot of many factors
that influenced the macroeconomic environment, for example, terms
of international trade, exchange rate volatility, domestic political
attitude. During the period from 2013 to 2015, Chinese economy slowed
down and it stirred a turmoil over the global commodity markets. Besides
the watches, other commodity importation shrank as well. While the
flagging economy would have weakened the imports of a myriad of commodities,
we employ PDA to control such overall effect in the hope to better
isolate the impact of the anti-corruption campaign.

\subsection{Results}

The dependent variable is set as the monthly growth rate of luxury
watch import in US dollars, and the independent variables are chosen
by the forward selection out of the import growth rates of 88 commodities.\footnote{To ensure that the control units are insusceptible to the anti-corruption
policy, 7 categories commonly consumed as bribe goods or conspicuous
consumption are excluded. These 7 categories are (with the UN Comtrade
Database code in the parenthesis): Beverages, spirits and vinegar
(22), Tobacco and manufactured tobacco substitutes (24), Essential
oils, perfumes, cosmetics, toiletries (33), Articles of leather, animal
gut, harness, travel goods (42), Fur-skins and artificial fur, manufactures
thereof (43), Pearls, precious stones, metals, coins, etc (71), Clocks
and watches and parts thereof (91) and Works of art, collectors pieces
and antiques (97). As a result, $88$ out of the total 95 categories
are left to serve as the pool of control units.} We use the growth rate instead of the level data to avoid time series
non-stationarity. January 2013 is regarded as the time of the treatment,
which is the month immediately after the \emph{Eight-Point Policy}
announcement. There are 35 pre-treatment observations ranging from
February 2010 to December 2012, and 36 post-treatment observations
spanning from January 2013 to December 2015. The same automated forward
selection algorithm as in the simulation chooses 3 control units.\footnote{The selected categories are ``knitted or crocheted fabric,'' ``cork
and articles of cork,'' and ``salt, sulfur, earth, stone, plaster,
lime and cement.''}

\begin{figure}
\begin{centering}
\includegraphics[scale=0.65]{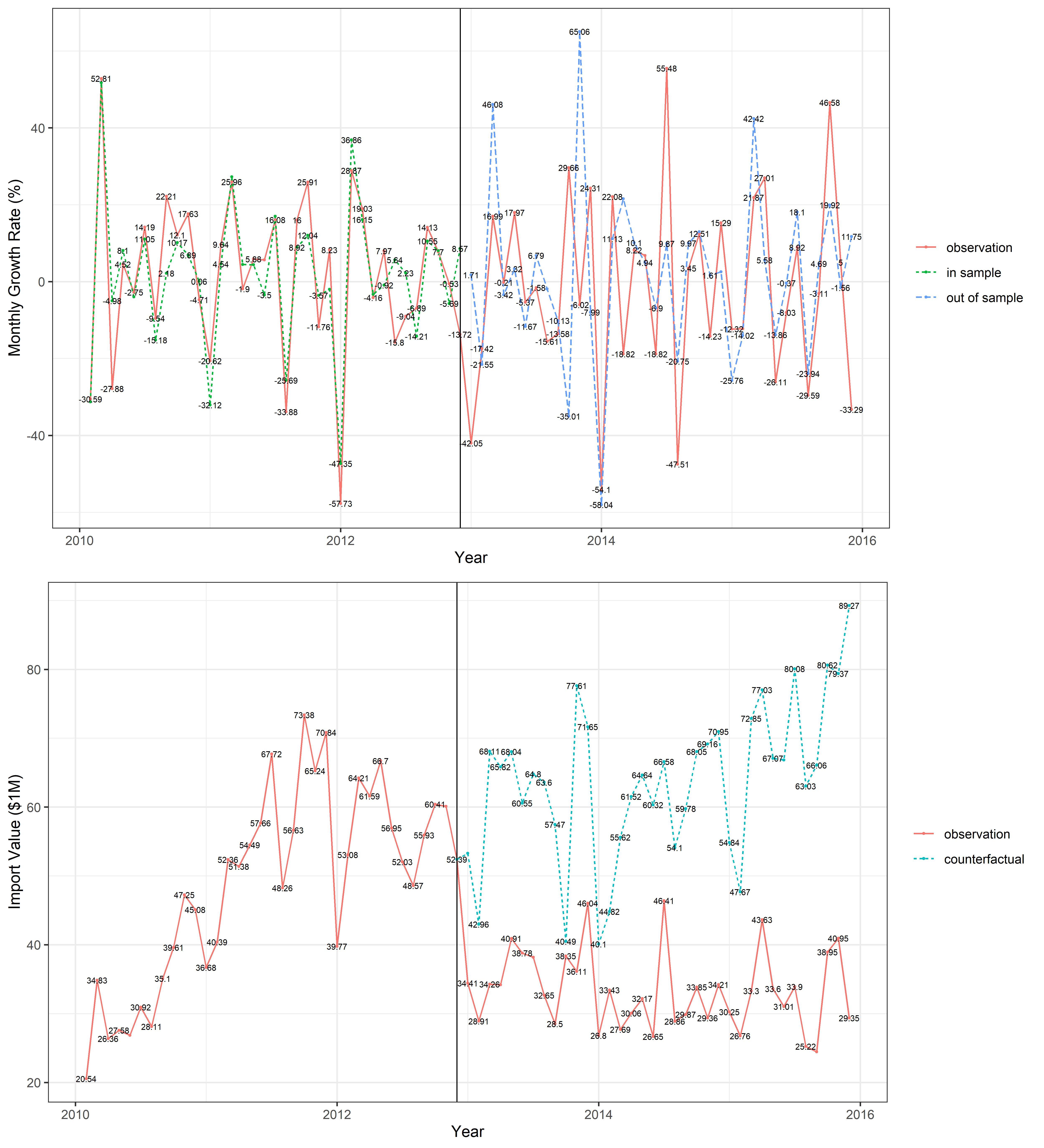}
\par\end{centering}
\begin{raggedright}
\small Note: The vertical line in the middle highlights the time
of the treatment January 2013. 
\par\end{raggedright}
\centering{}\caption{\label{fig:Luxury-Watches-Import:} Luxury Watches Import: Real Growth
and Counterfactual Prediction}
\end{figure}

With the estimated model, we predict the counterfactuals $\widehat{y}_{0t}^{0}$
and estimate treatment effect. Figure \ref{fig:Luxury-Watches-Import:}\textbf{
}displays the actual luxury watches import growth (solid line) and
its estimated counterparts without anti-corruption campaign (dashed
line). The upper subgraph shows the growth rate; the lower one shows
the value in US dollars, where the counterfactual in monetary value
is constructed according to the predicted growth rate. Before the
intervention, the model fits the real data quite well and the R-squared
of the selected model is 77.85\%. After January 2013, if the anti-corruption
policy had not been implemented, the import growth rate would have
followed the dashed line, which is visibly higher than the realizations.
In particular, in January 2013 the import value slumped by a whopping
42\%. In contrast, our counterfactual prediction suggests it would
have increased by 1.7\%. The ATE over the post-treatment period is
\[
\frac{1}{36}\sum_{t\in\mathcal{T}_{2}}\hat{\Delta}_{\widehat{U}_{\widehat{R}}t}=-3.09\%,
\]
which means that on average the anti-corruption campaign slowed down
the luxury watch import by 3.09\% per month. The $t$-statistic is
$-2.457$, with a $p$-value $1.40\%$. It rejects the null hypothesis
of zero ATE at 5\% size. Accumulating such a monthly ATE over 36 months
leads to roughly two thirds of reduction in importation ($\left(1-0.0309\right)^{36}=0.323$),
which is manifested in the lower subgraph. In December 2015, while
the realized import was 29.35 million US dollars, the counterfactual
predicts 89.27 million without the campaign. Our empirical evidence
suggests that China's anti-corruption has been effective in slashing
the luxury watch import.

\section{Conclusion}

In this paper, we propose using forward selection to choose control
units in PDA and then carrying out standard hypothesis testing. Forward
selection method is computationally much more efficient than the exhaustive
search for the best subset. We establish asymptotic theory for the
nearly optimality of forward selection, and show validity of conducting
post-selection inference for the ATE by the $t$-statistic based on
the selected set. Our theory is valid no matter the true coefficient
in the linear regression model is sparse or dense. These extensions
widen the applicability of PDA to real world high-dimensional problems
in modern data-rich environments.

\section*{Acknowledgements} 

Shi acknowledges the financial support from the Hong Kong Research
Grants Council No.24614817. We thank Cheng Hsiao, Qi Li, Peter Phillips,
Jeffrey Wooldridge and Yinchu Zhu for helpful comments, and Yishu
Wang for excellent research assistance. All remaining errors are ours.

\bigskip \bigskip \bigskip

%\onehalfspacing
%\normalsize
\setcounter{footnote}{0}
\setcounter{table}{0} 
\setcounter{figure}{0} 
\setcounter{equation}{0} 
\renewcommand{\thefootnote}{A\arabic{footnote}} 
\renewcommand{\theequation}{A\arabic{equation}} 
\renewcommand{\thefigure}{B\arabic{figure}} 
\renewcommand{\thetable}{B\arabic{table}} 

\appendix

\section{Technical Proofs}

\subsection{Proof of Lemma \ref{lem:pop=000026sam}}
\begin{proof}
Let \textbf{$\mathbf{y}_{j}:=(y_{jt})_{j\in\mathcal{T}_{1}}$} be
the $T_{1}\times1$ vector of time series for the $j$-th unit, and
let the $\text{\ensuremath{T_{1}\times\left|U\right|} matrix }\mathbf{Y}_{U}:=(\mathbf{y}_{j})_{j\in U}$,
where $U$ is a generic subset of $\mathcal{N}$. Let $\mathbf{P}_{U}:=\mathbf{Y}_{U}\left(\mathbf{Y}_{U}'\mathbf{Y}_{U}\right)^{-}\mathbf{Y}_{U}'$
be the projection matrix for the linear space spanned by $\mathbf{Y}_{U}$,
and $\mathbf{P}_{U}^{\perp}:=\mathbf{I}-\mathbf{P}_{U}$. 

\textbf{Part (a).} For any $U\subset\mathcal{N}$, whose cardinality
is $u=\left|U\right|$, define
\begin{align}
\hat{L}_{U}:=\frac{1}{T_{1}}\mathbf{y}_{0}'\mathbf{P}_{U}\mathbf{y}_{0} & =\frac{\mathbf{y}_{0}'\mathbf{Y}_{U}}{T_{1}}\left(\frac{\mathbf{Y}_{U}'\mathbf{Y}_{U}}{T_{1}}\right)^{-1}\frac{\mathbf{Y}_{U}'\mathbf{y}_{0}}{T_{1}}\nonumber \\
 & =\left(\mathcal{E}_{\left(1\right)}\left[\mathbf{y}_{Ut}y_{0t}\right]+\boldsymbol{\zeta}_{U}\right)'\left(\boldsymbol{\Sigma}_{U}+\mathbf{V}_{U}\right)^{-1}\left(\mathcal{E}_{\left(1\right)}\left[\mathbf{y}_{Ut}y_{0t}\right]+\boldsymbol{\zeta}_{U}\right)\label{eq:L_hat_U}
\end{align}
where $\boldsymbol{\zeta}_{U}:=\mathbb{E}_{\left(1\right)}\left[\mathbf{y}_{Ut}y_{0t}\right]-\mathcal{E}_{\left(1\right)}\left[\mathbf{y}_{Ut}y_{0t}\right]$,
$\boldsymbol{\Sigma}_{U}:=\mathcal{E}_{\left(1\right)}\left[\mathbf{y}_{Ut}\mathbf{y}_{Ut}'\right]$,
and $\mathbf{V}_{U}=\left(v_{ij}\right)_{i,j\in U}:=\mathbb{E}_{\left(1\right)}\left[\mathbf{y}_{Ut}\mathbf{y}_{Ut}'\right]-\boldsymbol{\Sigma}_{U}$.
Under Assumption \ref{assu:2nd-moment}(a), we have $\left\Vert \boldsymbol{\zeta}_{U}\right\Vert _{\infty}=O_{p}(\sqrt{\left(\log N\right)/T_{1}})$,
where $\left\Vert \cdot\right\Vert _{\infty}$ is the sup-norm of
a vector. The maximum eigenvalue of $\mathbf{V}_{U}$ is bounded by
\begin{equation}
\phi_{\max}\left(\mathbf{V}_{U}\right)\leq u\max_{i,j\in U}\left(v_{ij}\right)=O_{p}(u\sqrt{\left(\log N\right)/T_{1}})=o_{p}\left(1\right),\label{eq:eig_V}
\end{equation}
where the stochastic order again follows by Assumption \ref{assu:2nd-moment}(a).

Under Assumption \ref{assu:eigen}, (\ref{eq:eig_V}) implies that
when $N$ is sufficiently large the maximum eigenvalue of $\mathbf{V}_{U}$
will be dominated by the minimum eigenvalue of $\boldsymbol{\Sigma}_{U}$
with probability approaching one (w.p.a.1), and therefore 
\begin{align}
\left(\boldsymbol{\Sigma}_{U}+\mathbf{V}_{U}\right)^{-1} & =\boldsymbol{\Sigma}_{U}^{-1/2}\left(\mathbf{I}+\boldsymbol{\Sigma}_{U}^{1/2}\mathbf{V}_{U}\boldsymbol{\Sigma}_{U}^{1/2}\right)^{-1}\boldsymbol{\Sigma}_{U}^{-1/2}\nonumber \\
 & =\boldsymbol{\Sigma}_{U}^{-1/2}\big(\mathbf{I}-\sum_{l=1}^{\infty}(-\boldsymbol{\Sigma}_{U}^{1/2}\mathbf{V}_{U}\boldsymbol{\Sigma}_{U}^{1/2})^{l}\big)\boldsymbol{\Sigma}_{U}^{-1/2}=\boldsymbol{\Sigma}_{U}^{-1/2}\left(\mathbf{I}+\boldsymbol{\varXi}\right)\boldsymbol{\Sigma}_{U}^{-1/2},\label{eq:Sigma+V}
\end{align}
where $\boldsymbol{\varXi}:=\sum_{l=1}^{\infty}\left(-1\right)^{l+1}(\boldsymbol{\Sigma}_{U}^{1/2}\mathbf{V}_{U}\boldsymbol{\Sigma}_{U}^{1/2})^{l}$.
Assumption \ref{assu:eigen} also implies 
\begin{align*}
\phi_{\max}\left(\boldsymbol{\Sigma}_{U}^{1/2}\mathbf{V}_{U}\boldsymbol{\Sigma}_{U}^{1/2}\right) & \leq\phi_{\max}\left(\mathbf{V}_{U}\right)\phi_{\max}\left(\boldsymbol{\Sigma}_{U}\right)=\phi_{\max}\left(\mathbf{V}_{U}\right)\phi_{\min}^{-1}\left(\boldsymbol{\Sigma}_{U}\right)\\
 & =O_{p}\left(u\sqrt{\left(\log N\right)/T_{1}}\right)/c=O_{p}\left(u\sqrt{\left(\log N\right)/T_{1}}\right)
\end{align*}
w.p.a.1 when $N$ is sufficiently large and we have 
\begin{equation}
\phi_{\max}\left(\boldsymbol{\varXi}\right)\leq\frac{\phi_{\max}\left(\boldsymbol{\Sigma}_{U}^{1/2}\mathbf{V}_{U}\boldsymbol{\Sigma}_{U}^{1/2}\right)}{1-\phi_{\max}\left(\boldsymbol{\Sigma}_{U}^{1/2}\mathbf{V}_{U}\boldsymbol{\Sigma}_{U}^{1/2}\right)}=O_{p}\left(u\sqrt{\left(\log N\right)/T_{1}}\right).\label{eq:Sigma+V2}
\end{equation}
Substitute (\ref{eq:Sigma+V}) into (\ref{eq:L_hat_U}):
\begin{align}
\hat{L}_{U} & =\left(\mathcal{E}_{\left(1\right)}\left[\mathbf{y}_{Ut}y_{0t}\right]+\boldsymbol{\zeta}_{U}\right)'\boldsymbol{\Sigma}_{U}^{1/2}\left(\mathbf{I}+\boldsymbol{\varXi}\right)\boldsymbol{\Sigma}_{U}^{1/2}\left(\mathcal{E}_{\left(1\right)}\left[\mathbf{y}_{Ut}y_{0t}\right]+\boldsymbol{\zeta}_{U}\right)\nonumber \\
 & =\left(\mathcal{E}_{\left(1\right)}\left[\mathbf{y}_{Ut}y_{0t}\right]+\boldsymbol{\zeta}_{U}\right)'\boldsymbol{\Sigma}_{U}\left(\mathcal{E}_{\left(1\right)}\left[\mathbf{y}_{Ut}y_{0t}\right]+\boldsymbol{\zeta}_{U}\right)\cdot\left(1+O_{p}\left(u\sqrt{\left(\log N\right)/T_{1}}\right)\right)\label{eq:lem10}
\end{align}
in view of (\ref{eq:Sigma+V2}). Notice that 
\begin{eqnarray}
 &  & \left(\mathcal{E}_{\left(1\right)}\left[\mathbf{y}_{Ut}y_{0t}\right]+\boldsymbol{\zeta}_{U}\right)'\boldsymbol{\Sigma}_{U}\left(\mathcal{E}_{\left(1\right)}\left[\mathbf{y}_{Ut}y_{0t}\right]+\boldsymbol{\zeta}_{U}\right)\nonumber \\
 & = & \mathcal{E}_{\left(1\right)}\left[\mathbf{y}_{Ut}y_{0t}\right]'\boldsymbol{\Sigma}_{U}\mathcal{E}_{\left(1\right)}\left[\mathbf{y}_{Ut}y_{0t}\right]+2\boldsymbol{\zeta}_{U}'\boldsymbol{\Sigma}_{U}\mathcal{E}_{\left(1\right)}\left[\mathbf{y}_{Ut}y_{0t}\right]+\boldsymbol{\zeta}_{U}'\boldsymbol{\Sigma}_{U}\boldsymbol{\zeta}_{U}\nonumber \\
 & = & L_{U}+2\boldsymbol{\zeta}_{U}'\boldsymbol{\Sigma}_{U}\mathcal{E}_{\left(1\right)}\left[\mathbf{y}_{Ut}y_{0t}\right]+\boldsymbol{\zeta}_{U}'\boldsymbol{\Sigma}_{U}\boldsymbol{\zeta}_{U},\label{eq:lem10_1}
\end{eqnarray}
where $L_{U}:=\mathcal{E}_{\left(1\right)}\left[\mathbf{y}_{Ut}y_{0t}\right]'\boldsymbol{\Sigma}_{U}\mathcal{E}_{\left(1\right)}\left[\mathbf{y}_{Ut}y_{0t}\right].$
The third term on the right-hand side of the above equation is bounded
by
\begin{equation}
\boldsymbol{\zeta}_{U}'\boldsymbol{\Sigma}_{U}\boldsymbol{\zeta}_{U}\leq\phi_{\min}^{-1}\left(\boldsymbol{\Sigma}_{U}\right)\left\Vert \boldsymbol{\zeta}_{U}\right\Vert _{2}^{2}\leq c^{-1}u\left\Vert \boldsymbol{\zeta}_{U}\right\Vert _{\infty}^{2}=O_{p}\left(u\left(\log N\right)/T_{1}\right),\label{eq:lem11}
\end{equation}
and the second term is bounded by 
\begin{align}
2\boldsymbol{\zeta}_{U}'\boldsymbol{\Sigma}_{U}\mathcal{E}_{\left(1\right)}\left[\mathbf{y}_{Ut}y_{0t}\right] & =2\left(\boldsymbol{\Sigma}_{U}^{1/2}\boldsymbol{\zeta}_{U}\right)'\left(\boldsymbol{\Sigma}_{U}^{1/2}\mathcal{E}_{\left(1\right)}\left[\mathbf{y}_{Ut}y_{0t}\right]\right)\leq2\left(\boldsymbol{\zeta}_{U}'\boldsymbol{\Sigma}_{U}\boldsymbol{\zeta}_{U}\right)^{1/2}\sqrt{L_{U}}\nonumber \\
 & \leq2\phi_{\min}^{-1/2}\left(\boldsymbol{\Sigma}_{U}\right)\cdot\left\Vert \boldsymbol{\zeta}_{U}\right\Vert _{2}\cdot\sqrt{L_{U}}\leq2c^{-1/2}\cdot\sqrt{u}\left\Vert \boldsymbol{\zeta}_{U}\right\Vert _{\infty}\cdot\sqrt{L_{U}}\nonumber \\
 & =O_{p}\left(\sqrt{u\left(\log N\right)/T_{1}}\right),\label{eq:lem12}
\end{align}
where the first inequality follows by the Cauchy-Schwarz inequality.
Substituting (\ref{eq:lem10_1}), (\ref{eq:lem11}) and (\ref{eq:lem12})
into (\ref{eq:lem10}) gives
\begin{align}
\hat{L}_{U} & =\left(L_{U}+O_{p}\left(\sqrt{u\frac{\log N}{T_{1}}}\right)\right)\left(1+O_{p}\left(\sqrt{u\frac{\log N}{T_{1}}}\right)\right)=L_{U}+O_{p}\left(\sqrt{u\frac{\log N}{T_{1}}}\right).\label{eq:L_hat}
\end{align}

Finally, when $U=\emptyset$ we denote $\widehat{\sigma}_{\emptyset}^{2}$
as the sample variance of $\left\{ y_{0t}\right\} _{t\in\mathcal{T}_{1}}$
when no regressors are considered, and correspondingly $\sigma_{\emptyset}^{2}:=\mathcal{E}_{\left(1\right)}\left[y_{0t}^{2}\right]$.
Obviously, $\widehat{\sigma}_{\emptyset}^{2}-\sigma_{\emptyset}^{2}=O_{p}\left(T_{1}^{-1}\right)$.
By definition $\widehat{L}_{U}=\widehat{\sigma}_{\emptyset}^{2}-\widehat{\sigma}_{U}^{2}$
and $L_{U}=\sigma_{\emptyset}^{2}-\sigma_{U}^{2}$, and it follows
\[
\widehat{\sigma}_{U}^{2}-\sigma_{U}^{2}=\left(\widehat{\sigma}_{\emptyset}^{2}-\sigma_{\emptyset}^{2}\right)-(\widehat{L}_{U}-L_{U}).
\]
 Since the above equality (\ref{eq:L_hat}) holds uniformly for all
$U$ and Assumption \ref{assu:eigen} is stated for $R$, we have
\[
\max_{\mathcal{U}_{(1+\delta_{1})R}}\left|\widehat{L}_{U}-L_{U}\right|=O_{p}\left(\sqrt{\left(1+\delta_{1}\right)R\left(\log N\right)/T_{1}}\right)=O_{p}\left(\sqrt{R\left(\log N\right)/T_{1}}\right)
\]
 for $\delta_{1}$ is a universal constant. Part (a) follows in view
of the last two display expressions. 

\textbf{Part (b).} For any $U\in\mathcal{U}_{(1+\delta_{1})R}$, the
orthogonality between $\mathbf{y}_{Ut}$ and $\varepsilon_{Ut}$ implies
\begin{align*}
\mathcal{E}_{\left(1\right)}\left[y_{0t}^{2}\right] & =\mathcal{E}_{\left(1\right)}\left[\left(\mathbf{y}_{Ut}'\boldsymbol{\beta}_{U}^{0}+\varepsilon_{Ut}\right)^{2}\right]=\boldsymbol{\beta}_{U}^{0\prime}\mathcal{E}_{\left(1\right)}\left[\mathbf{y}_{Ut}\mathbf{y}_{Ut}'\right]\boldsymbol{\beta}_{U}^{0}+\mathcal{E}_{\left(1\right)}\left[\varepsilon_{Ut}^{2}\right]\\
 & \geq\boldsymbol{\beta}_{U}^{0\prime}\mathcal{E}_{\left(1\right)}\left[\mathbf{y}_{Ut}\mathbf{y}_{Ut}'\right]\boldsymbol{\beta}_{U}^{0}\geq\left\Vert \boldsymbol{\beta}_{U}^{0}\right\Vert _{2}^{2}\eta_{u}.
\end{align*}
Therefore under Assumptions \ref{assu:eigen} and \ref{assu:2nd-moment}(b)
when $N$ is sufficiently large,
\begin{equation}
\max_{\mathcal{U}_{(1+\delta_{1})R}}\left\Vert \boldsymbol{\beta}_{U}^{0}\right\Vert _{2}\leq\sqrt{\mathcal{E}_{\left(1\right)}\left[y_{0t}^{2}\right]/\eta_{\left(1+\delta_{1}\right)R}}\leq\sqrt{C/c}<\infty\label{eq:beta_bound}
\end{equation}
so that the population coefficients are bounded from above. The OLS
estimator can be written in the closed form
\begin{align*}
\widehat{\boldsymbol{\beta}}_{U} & =\left(\mathbf{Y}_{U}'\mathbf{Y}_{U}/T_{1}\right)^{-1}\left(\mathbf{Y}_{U}^{\prime}\mathbf{y}_{0}/T_{1}\right)=\left(\boldsymbol{\Sigma}_{U}+\mathbf{V}_{U}\right)^{-1}\left(\mathcal{E}_{\left(1\right)}\left[\mathbf{y}_{Ut}y_{0t}\right]+\boldsymbol{\zeta}_{U}\right)\\
 & =\boldsymbol{\Sigma}_{U}^{-1/2}\left(\mathbf{I}+\boldsymbol{\varXi}\right)\boldsymbol{\Sigma}_{U}^{-1/2}\left(\mathcal{E}_{\left(1\right)}\left[\mathbf{y}_{Ut}y_{0t}\right]+\boldsymbol{\zeta}_{U}\right)\\
 & =\boldsymbol{\Sigma}_{U}^{-1}\mathcal{E}_{\left(1\right)}\left[\mathbf{y}_{Ut}y_{0t}\right]+\boldsymbol{\Sigma}_{U}^{-1/2}\boldsymbol{\varXi}\boldsymbol{\Sigma}_{U}^{-1/2}\left(\mathcal{E}_{\left(1\right)}\left[\mathbf{y}_{Ut}y_{0t}\right]+\boldsymbol{\zeta}_{U}\right)+\boldsymbol{\Sigma}_{U}^{-1}\boldsymbol{\zeta}_{U}\\
 & =\boldsymbol{\beta}_{U}^{0}+\boldsymbol{\Sigma}_{U}^{-1/2}\boldsymbol{\varXi}\boldsymbol{\Sigma}_{U}^{1/2}\boldsymbol{\beta}_{U}^{0}+\boldsymbol{\Sigma}_{U}^{-1/2}\boldsymbol{\varXi}\boldsymbol{\Sigma}_{U}^{1/2}\boldsymbol{\Sigma}_{U}^{-1}\boldsymbol{\zeta}_{U}+\boldsymbol{\Sigma}_{U}^{-1}\boldsymbol{\zeta}_{U}
\end{align*}
in view of $\boldsymbol{\beta}_{U}^{0}=\boldsymbol{\Sigma}_{U}^{-1}\mathcal{E}_{\left(1\right)}\left[\mathbf{y}_{Ut}y_{0t}\right]$.
Subtract $\boldsymbol{\beta}_{U}^{0}$ and take the $\left\Vert \cdot\right\Vert _{2}$-norm
on both sides of the above equation: 
\begin{align*}
\left\Vert \widehat{\boldsymbol{\beta}}_{U}-\boldsymbol{\beta}_{U}^{0}\right\Vert _{2} & \leq\left\Vert \boldsymbol{\Sigma}_{U}^{-1/2}\boldsymbol{\varXi}\boldsymbol{\Sigma}_{U}^{1/2}\boldsymbol{\beta}_{U}^{0}\right\Vert _{2}+\left\Vert (\boldsymbol{\Sigma}_{U}^{-1/2}\boldsymbol{\varXi}\boldsymbol{\Sigma}_{U}^{1/2}+\mathbf{I})\boldsymbol{\Sigma}_{U}^{-1}\boldsymbol{\zeta}_{U}\right\Vert _{2}\\
 & \leq\phi_{\max}\left(\boldsymbol{\Sigma}_{U}^{-1/2}\boldsymbol{\varXi}\boldsymbol{\Sigma}_{U}^{1/2}\right)\left\Vert \boldsymbol{\beta}_{U}^{0}\right\Vert _{2}+\left(\phi_{\max}\left(\boldsymbol{\Sigma}_{U}^{-1/2}\boldsymbol{\varXi}\boldsymbol{\Sigma}_{U}^{1/2}\right)+1\right)\left\Vert \boldsymbol{\Sigma}_{U}^{-1}\boldsymbol{\zeta}_{U}\right\Vert _{2}\\
 & \leq\phi_{\max}\left(\boldsymbol{\varXi}\right)\left\Vert \boldsymbol{\beta}_{U}^{0}\right\Vert _{2}+\left(\phi_{\max}\left(\boldsymbol{\varXi}\right)+1\right)\phi_{\max}\left(\boldsymbol{\Sigma}_{U}^{-1}\right)\left\Vert \boldsymbol{\zeta}_{U}\right\Vert _{2}\\
 & =O_{p}\left(u\sqrt{\left(\log N\right)/T_{1}}\right)\left\Vert \boldsymbol{\beta}_{U}^{0}\right\Vert _{2}+\left(O_{p}\left(u\sqrt{\left(\log N\right)/T_{1}}\right)+1\right)\eta_{u}^{-1}\sqrt{u}\left\Vert \boldsymbol{\zeta}_{U}\right\Vert _{\infty}\\
 & =O_{p}\left(u\sqrt{\left(\log N\right)/T_{1}}\right)+O_{p}\left(u^{\frac{3}{2}}\sqrt{\left(\log N\right)/T_{1}}\right)=O_{p}\left(\sqrt{R^{3}\left(\log N\right)/T_{1}}\right),
\end{align*}
where the first equality follows by (\ref{eq:Sigma+V2}), and the
second equality by (\ref{eq:beta_bound}) as well as Assumptions \ref{assu:eigen}
and \ref{assu:2nd-moment}(a).
\end{proof}

\subsection{Proof of Theorem \ref{thm:uniform_inference}}

Since we consider the set of DGPs $\mathcal{M}$ which satisfy Assumptions
\ref{assu:eigen}, \ref{assu:2nd-moment}, \ref{assu:inference} and
\ref{assu:alpha_phi_coef} uniformly, the stochastic orders in Lemma
\ref{lem:pop=000026sam} are uniform over $\mathcal{M}.$ Moreover,
all stochastic orders in this section are uniform over $\mathcal{M}$
as well.

Given an index set $U\subset\mathcal{N},$ define a $t$-statistic
$\mathcal{Z}_{U}^{*}:=\rho_{U}^{*-1}\sqrt{T_{2}}\mathbb{E}_{\left(2\right)}[\varepsilon_{Ut}],$
where $\rho_{U}^{*2}:=T_{2}^{-1}E[(\sum_{t\in\mathcal{T}_{2}}\varepsilon_{Ut})^{2}]$.
Next, denote its truncated version with the data in $t\in\left\{ k+1,k+2\ldots,T_{2}\right\} $
as $\mathcal{Z}_{U}^{\left(k\right)*}:=(\rho_{U}^{\left(k\right)*}\sqrt{T_{2}-k})^{-1}\sum_{t=k+1}^{T_{2}}\varepsilon_{Ut},$
where $\rho_{U}^{\left(k\right)*2}:=(T_{2}-k)^{-1}E[(\sum_{t=k+1}^{T_{2}}\varepsilon_{Ut})^{2}]$.
The truncated version $\mathcal{Z}_{U}^{\left(k\right)*}$ drops the
observations during $t\in\left\{ 1,\ldots,k\right\} $ \textemdash{}
those in $\mathcal{T}_{2}$ but are close to the end of $\mathcal{T}_{1}$.
All the $t$-statistics in this paper depend on the DGP $M\in\mathcal{M}$
while we suppress ``$M$'' for concise notation when no confusion
arises.
\begin{lem}
\label{lem:converge_z} Suppose the Assumptions \ref{assu:eigen},
\ref{assu:2nd-moment}, \ref{assu:inference} and \ref{assu:alpha_phi_coef}
hold and the null hypothesis is $\mathbb{H}_{0}$ is true.
\begin{enumerate}
\item If $T_{1}^{-1}R^{4}\log^{2}T_{2}\log^{2}N\to0$ and $1/\tau+\tau/\log T_{2}\to0$
as $N\to\infty$, then 
\[
\sup_{M\in\mathcal{M}}\text{\ensuremath{\max}}_{\mathcal{U}_{R}}\left|\mathcal{Z}_{U}-\mathcal{Z}_{U}^{*}\right|\stackrel{p}{\to}0.
\]
\item If $k=k\left(N\right)\to\infty$ and $k/T_{2}\to0$ as $N\to\infty$,
then 
\[
\sup_{M\in\mathcal{M}}\text{\ensuremath{\max}}_{\mathcal{U}_{R}}\left|\mathcal{Z}_{U}^{*}-\mathcal{Z}_{U}^{\left(k\right)*}\right|\stackrel{p}{\to}0.
\]
\end{enumerate}
\end{lem}
\begin{rem}
Lemma \ref{lem:converge_z}(a) is about the uniform asymptotic equivalence
between $\mathcal{Z}_{U}$ and $\mathcal{Z}_{U}^{*}$ under the null,
which means that the former has the same asymptotic distribution as
the latter. For the latter is a statistic involving no estimated parameters
from the pre-treatment data, it is much easier to pin down its asymptotic
distribution by borrowing convergence in distribution results from
the literature of probability theory. Result (b) is about the uniform
asymptotic equivalence between $\mathcal{Z}_{U}^{*}$ and $\mathcal{Z}_{U}^{\left(k\right)*}$.
Due to weak dependence, as $k$ gets larger $\mathcal{Z}_{U}^{\left(k\right)*}$
is asymptotically independent of the pre-treatment data.
\end{rem}
\begin{proof}[Proof of Lemma \ref{lem:converge_z}]
 \textbf{Part (a)}. We introduce a new $t$-statistic $\mathcal{Z}_{\tau U}^{*}$
which serves as a bridge to connect $\mathcal{Z}_{U}$ and $\mathcal{Z}_{U}^{*}$.
Let 
\[
\mathcal{Z}_{\tau U}^{*}:=\widehat{\rho}_{\tau U}^{*-1}\sqrt{T_{2}}\mathbb{E}_{\left(2\right)}\left[\varepsilon_{Ut}\right]=\widehat{\rho}_{\tau U}^{*-1}\sqrt{T_{2}}\mathbb{E}_{\left(2\right)}\left[y_{0t}-\mathbf{y}_{Ut}'\boldsymbol{\beta}_{U}^{0}\right].
\]
 This $\mathcal{Z}_{\tau U}^{*}$ is an infeasible version of $\mathcal{Z}_{U}$
as if the true coefficient $\boldsymbol{\beta}_{U}^{0}$ is known,
and the estimated long-run variance $\widehat{\rho}_{\tau U}^{*2}:=T_{2}^{-1}\sum_{t,s\in\mathcal{T}_{2},\left|t-s\right|\leq\tau}\varepsilon_{Ut}\varepsilon_{Us}$
is the infeasible counterpart of $\widehat{\rho}_{\tau U}$ with known
$\boldsymbol{\beta}_{U}^{0}$.

Since $\Delta_{t}$ only changes the mean, all distributional changes
are absorbed by $(\varepsilon_{t})_{t\in\mathcal{T}_{2}}$. Under
the null hypothesis $\mathbb{H}_{0}$, we replace $\widehat{\Delta}_{Ut}$
by $\widehat{\varepsilon}_{Ut}$ in the computation. Uniformly for
all index sets $\mathcal{U}_{R}$ the difference between the nominators
of $\mathcal{Z}_{\tau U}^{*}$ and $\mathcal{Z}_{U}$ is bounded by
\begin{eqnarray}
 &  & \bigg|\sqrt{T_{2}}\mathbb{E}_{\left(2\right)}\left[\widehat{\varepsilon}_{Ut}-\varepsilon_{Ut}\right]\bigg|=\bigg|(\widehat{\boldsymbol{\beta}}_{U}-\boldsymbol{\beta}_{U}^{0})'\sqrt{T_{2}}\mathbb{E}_{\left(2\right)}\left[\mathbf{y}_{Ut}\right]\bigg|\leq\big\Vert\widehat{\boldsymbol{\beta}}_{U}-\boldsymbol{\beta}_{U}^{0}\big\Vert_{2}\big\Vert\sqrt{T_{2}}\mathbb{E}_{\left(2\right)}\left[\mathbf{y}_{Ut}\right]\big\Vert_{2}\nonumber \\
 & \leq & \big\Vert\widehat{\boldsymbol{\beta}}_{U}-\boldsymbol{\beta}_{U}^{0}\big\Vert_{2}\sqrt{R}\cdot\sqrt{T_{2}}\max_{j\in U}\left|\mathbb{E}_{\left(2\right)}\left[y_{it}\right]\right|=O_{p}\left(\sqrt{R^{3}\left(\log N\right)/T_{1}}\right)\sqrt{R}O_{p}\left(\sqrt{\log N}\right)\nonumber \\
 & = & O_{p}\left(\sqrt{R^{4}\left(\log^{2}N\right)/T}\right),\label{eq:lrvar_num}
\end{eqnarray}
where the first inequality follows by the Cauchy-Schwarz inequality,
and the stochastic order by Assumption \ref{assu:inference}(a).

The difference between the long-run variances is bounded by 
\begin{eqnarray}
 &  & \left|\widehat{\rho}_{\tau U}^{*2}-\widehat{\rho}_{\tau U}^{2}\right|=T_{2}^{-1}\bigg|\sum_{t,s\in\mathcal{T}_{2},\left|t-s\right|\leq\tau}(\widehat{\varepsilon}_{Ut}\widehat{\varepsilon}_{Us}-\varepsilon_{Ut}\varepsilon_{Us})\bigg|\nonumber \\
 & = & T_{2}^{-1}\bigg|\sum_{t,s\in\mathcal{T}_{2},\left|t-s\right|\leq\tau}\left\{ \left(\varepsilon_{Ut}-\mathbf{y}_{Ut}'(\widehat{\boldsymbol{\beta}}_{U}-\boldsymbol{\beta}_{U}^{0})\right)\left(\varepsilon_{Us}-\mathbf{y}_{Us}'(\widehat{\boldsymbol{\beta}}_{U}-\boldsymbol{\beta}_{U}^{0})\right)-\varepsilon_{Ut}\varepsilon_{Us}\right\} \bigg|\nonumber \\
 & \leq & (\widehat{\boldsymbol{\beta}}_{U}-\boldsymbol{\beta}_{U}^{0})'T_{2}^{-1}\bigg|\sum_{t,s\in\mathcal{T}_{2},\left|t-s\right|\leq\tau}\mathbf{y}_{Ut}\mathbf{y}_{Us}^{\prime}\bigg|(\widehat{\boldsymbol{\beta}}_{U}-\boldsymbol{\beta}_{U}^{0})+2T_{2}^{-1}(\widehat{\boldsymbol{\beta}}_{U}-\boldsymbol{\beta}_{U}^{0})'\bigg|\sum_{t,s\in\mathcal{T}_{2},\left|t-s\right|\leq\tau}\mathbf{y}_{Ut}\varepsilon_{Us}\bigg|\nonumber \\
 & \leq & \Vert\widehat{\boldsymbol{\beta}}_{U}-\boldsymbol{\beta}_{U}^{0}\Vert_{2}^{2}\cdot\tau\phi_{\max}\big(\mathbb{E}_{\left(2\right)}[\mathbf{y}_{Ut}\mathbf{y}_{Ut}^{\prime}]\big)\nonumber \\
 &  & +2\Vert\widehat{\boldsymbol{\beta}}_{U}-\boldsymbol{\beta}_{U}^{0}\Vert_{2}\cdot\tau\max_{0\leq l\leq\tau}\big\Vert\mathbb{E}_{\left(2\right)}[\mathbf{y}_{Ut}\varepsilon_{U,t+l}]\cdot\boldsymbol{1}\left\{ 1\leq t+l\leq T_{2}\right\} \big\Vert_{2}\label{eq:lrvar}
\end{eqnarray}
by the triangle inequality. In the above inequality (\ref{eq:lrvar}),
we have 
\begin{align}
\phi_{\max}\big(\mathbb{E}_{\left(2\right)}[\mathbf{y}_{Ut}\mathbf{y}_{Ut}^{\prime}]\big) & \leq u\max_{j\in\mathcal{N}}\mathbb{E}_{\left(2\right)}[y_{jt}^{2}]=u\left(\max_{j\in\mathcal{N}}\mathcal{E}_{\left(2\right)}\left[y_{jt}^{2}\right]+o_{p}\left(1\right)\right)=O_{p}\left(R\right)\label{eq:lrvar1}
\end{align}
 by Assumption \ref{assu:inference}(b) and (c). Similarly, the cross
term in the right-hand side of (\ref{eq:lrvar}) is bounded by 
\begin{eqnarray}
 &  & \max_{0\leq l\leq\tau}\big\Vert\mathbb{E}_{\left(2\right)}[\mathbf{y}_{Ut}\varepsilon_{U,t+l}]\cdot\boldsymbol{1}\left\{ 1\leq t+l\leq T_{2}\right\} \big\Vert_{2}\nonumber \\
 & \leq & \sqrt{u}\max_{0\leq l\leq\tau}\max_{j\in U}|\mathbb{E}_{\left(2\right)}[y_{jt}\varepsilon_{U,t+l}\cdot\boldsymbol{1}\left\{ 1\leq t+l\leq T_{2}\right\} ]|\leq\sqrt{u}\max_{j\in U}\big(\mathbb{E}_{\left(2\right)}[y_{jt}^{2}]\mathbb{E}_{\left(2\right)}[\varepsilon_{Ut}^{2}]\big)^{1/2}\nonumber \\
 & \leq & \sqrt{u}\max_{j\in U}\big(\mathbb{E}_{\left(2\right)}[y_{jt}^{2}]\mathbb{E}_{\left(2\right)}[y_{0t}^{2}]\big)^{1/2}\leq\sqrt{u}\max_{j\in\mathcal{N}_{0}}\mathbb{E}_{\left(2\right)}[y_{jt}^{2}]\nonumber \\
 & = & \sqrt{u}\left(\max_{j\in\mathcal{N}_{0}}\mathcal{E}_{\left(2\right)}\left[y_{jt}^{2}\right]+o_{p}\left(1\right)\right)=O_{p}(R^{1/2})\label{eq:lrvar2}
\end{eqnarray}
where the first and the second inequality follow by the Cauchy-Schwarz
inequality. 

Notice that (\ref{eq:lrvar1}) and (\ref{eq:lrvar2}) hold uniformly
over $\mathcal{U}_{R}$. Substitute (\ref{eq:lrvar1}), (\ref{eq:lrvar2})
and Lemma \ref{lem:pop=000026sam}(b) into (\ref{eq:lrvar}), and
notice $\tau/\log T_{2}\to0$, we have 
\begin{align*}
\text{\ensuremath{\max}}_{\mathcal{U}_{R}}\left|\widehat{\rho}_{\tau U}^{*2}-\widehat{\rho}_{\tau U}^{2}\right| & \leq\tau O_{p}\left(R^{3}\left(\log N\right)/T_{1}\right)O_{p}\left(R\right)+\tau O_{p}\left(\sqrt{R^{3}\left(\log N\right)/T_{1}}\right)O_{p}(R^{1/2})\\
 & =O_{p}\left(\tau\sqrt{R^{4}\left(\log N\right)/T_{1}}\right)=O_{p}\left(\sqrt{R^{4}\log^{2}T_{2}\left(\log N\right)/T_{1}}\right)
\end{align*}
The above inequality, along with the boundedness of the population
long-run variance as in Assumptions \ref{assu:inference}(d) and (e),
ensures that the estimation error in the denominator is asymptotically
negligible under the rate condition $R^{4}\log^{2}T_{2}\left(\log N\right)/T_{1}\to0$.
In other words, the stochastic order of the difference between $\mathcal{Z}_{\tau U}^{*}$
and $\mathcal{Z}_{U}$ is governed by the difference in the numerators
as in (\ref{eq:lrvar_num}). We have shown the asymptotic equivalence
$\text{\ensuremath{\max}}_{\mathcal{U}_{R}}\left|\mathcal{Z}_{U}-\mathcal{Z}_{\tau U}^{*}\right|=o_{p}\left(1\right)$. 

The nominators of $\mathcal{Z}_{\tau U}^{*}$ and $\mathcal{Z}_{U}^{*}$
are the same. Their denominators $\text{\ensuremath{\max}}_{\mathcal{U}_{R}}|\widehat{\rho}_{\tau U}^{*2}-\rho_{U}^{*2}|=o_{p}\left(1\right)$
since $\widehat{\rho}_{\tau U}^{*2}$ consistently estimates the long-run
variance $\rho_{U}^{*2}$, which is bounded above for all $\mathcal{U}_{R}$
according to Assumption \ref{assu:inference}(e). We thus have the
asymptotic equivalence $\text{\ensuremath{\max}}_{\mathcal{U}_{R}}\left|\mathcal{Z}_{\tau U}^{*}-\mathcal{Z}_{U}^{*}\right|=o_{p}\left(1\right)$.
Up to now we have $\text{\ensuremath{\max}}_{\mathcal{U}_{R}}\left|\mathcal{Z}_{U}-\mathcal{Z}_{U}^{*}\right|=o_{p}\left(1\right)$.
As the stochastic orders hold uniformly for all $M\in\mathcal{M}$,
we have establish part (a).

\textbf{Part (b)}. Regarding $\mathcal{Z}_{U}^{*}$ and its truncated
version $\mathcal{Z}_{U}^{\left(k\right)*}$, notice that the difference
of the nominators is
\begin{align}
\frac{1}{\sqrt{T_{2}}}\sum_{t\in\mathcal{T}_{2}}\varepsilon_{Ut}-\frac{1}{\sqrt{T_{2}-k}}\sum_{t=k+1}^{T_{2}}\varepsilon_{Ut} & =\frac{1}{\sqrt{T_{2}}}\sum_{t=1}^{k}\varepsilon_{Ut}-\left(\frac{1}{\sqrt{T_{2}-k}}-\frac{1}{\sqrt{T_{2}}}\right)\sum_{t=k+1}^{T_{2}}\varepsilon_{Ut}\nonumber \\
 & =:A_{1}+A_{2}\label{eq:partb_1}
\end{align}
The variance of the first term $A_{1}$ on the right-hand side of
(\ref{eq:partb_1}) is bounded by 
\begin{equation}
T_{2}^{-1}E\left[\big(\sum_{t=1}^{k}\varepsilon_{Ut}\big)^{2}\right]=\frac{k}{T_{2}}\cdot\frac{1}{k}E\left[\sum_{1\leq t,s\leq k}\varepsilon_{Ut}\varepsilon_{Us}\right]\leq\frac{k}{T_{2}}C\label{eq:partb_2}
\end{equation}
by Assumption \ref{assu:inference}(e), and therefore $A_{1}=O_{p}\left(\sqrt{k/T_{2}}\right)$.
Similarly, 
\begin{align*}
A_{2} & =\left(1-\sqrt{1-k/T_{2}}\right)\left(T_{2}-k\right)^{-1/2}\sum_{t=k+1}^{T_{2}}\varepsilon_{Ut}\\
 & =\frac{1}{2}\frac{k}{T_{2}}\left(1+o\left(1\right)\right)\cdot\left(T_{2}-k\right)^{-1/2}\sum_{t=k+1}^{T_{2}}\varepsilon_{Ut}=O_{p}\left(\sqrt{k/T_{2}}\right),
\end{align*}
where in the second equality we use the Taylor expansion $\sqrt{1-x}=1-0.5x+o\left(x\right)$
as $x\to0$ for approximation, and the stochastic order by the same
reasoning as in (\ref{eq:partb_2}). The orders of $A_{1}$ and $A_{2}$
ensure that the right-hand side of (\ref{eq:partb_1}) is $O_{p}(\sqrt{k/T_{2}})$
uniformly over $\mathcal{U}_{R}$.

Now we turn to the denominators. The difference in the population
long-run variance is 
\begin{align*}
\rho_{U}^{*2}-\rho_{U}^{\left(k\right)*2} & =T_{2}^{-1}E\left[(\sum_{t=1}^{k}\varepsilon_{Ut}+\sum_{t=k+1}^{T_{2}}\varepsilon_{Ut})^{2}\right]-(T_{2}-k)^{-1}E\left[(\sum_{t=k+1}^{T_{2}}\varepsilon_{Ut})^{2}\right]\\
 & =T_{2}^{-1}E\left[(\sum_{t=1}^{k}\varepsilon_{Ut})^{2}\right]+2T_{2}^{-1}E\left[\sum_{t=1}^{k}\varepsilon_{Ut}\sum_{t=k+1}^{T_{2}}\varepsilon_{Ut}\right]+\left(T_{2}^{-1}-(T_{2}-k)^{-1}\right)E\left[(\sum_{t=k+1}^{T_{2}}\varepsilon_{Ut})^{2}\right]\\
 & =:A_{3}+2A_{4}+A_{5}.
\end{align*}
Again by Assumption \ref{assu:inference}(e), using similar derivation
as in (\ref{eq:partb_2}) we have the first term $A_{3}=(k/T_{2})k^{-1}E[(\sum_{t=1}^{k}\varepsilon_{Ut})^{2}]=O(k/T_{2})$
and the third term $A_{5}=\left(T_{2}-k\right)^{-1}(k/T_{2})E[(\sum_{t=k+1}^{T_{2}}\varepsilon_{Ut})^{2}]=O\left(k/T_{2}\right)$.
The triangle inequality bounds the half of the second term: 
\begin{align*}
\left|A_{4}\right| & =T_{2}^{-1}\left|E\left[\sum_{t=1}^{k}\varepsilon_{Ut}\sum_{t=k+1}^{T_{2}}\varepsilon_{Ut}\right]\right|=T_{2}^{-1}\left|E\left[\sum_{t=1}^{k}\varepsilon_{Ut}(\sum_{t=k+1}^{2k}\varepsilon_{Ut}+\sum_{t=2k+1}^{T_{2}}\varepsilon_{Ut})\right]\right|\\
 & \leq T_{2}^{-1}\left|E\left[\sum_{t=1}^{k}\varepsilon_{Ut}\sum_{t=k+1}^{2k}\varepsilon_{Ut}\right]\right|+T_{2}^{-1}\left|E\left[\sum_{t=1}^{k}\varepsilon_{Ut}\sum_{t=2k+1}^{T_{2}}\varepsilon_{Ut}\right]\right|=:A_{6}+A_{7}.
\end{align*}
 The Cauchy-Schwarz inequality and Assumption \ref{assu:inference}(e)
imply
\[
A_{6}\leq\frac{k}{T_{2}}\sqrt{\frac{1}{k}E\left[\left(\sum_{t=1}^{k}\varepsilon_{Ut}\right)^{2}\right]\cdot\frac{1}{k}E\left[\left(\sum_{t=k+1}^{2k}\varepsilon_{Ut}\right)^{2}\right]}=O\left(k/T_{2}\right).
\]
Moreover, the two separated time series segments $\left(\varepsilon_{Ut}\right)_{t=1}^{k}$
and $\left(\varepsilon_{Ut}\right)_{t=2k+1}^{T_{2}}$ are asymptotically
independent under the $\phi$-mixing condition in Assumption \ref{assu:alpha_phi_coef},
and so are the empirical processes $k^{-1/2}\sum_{t=1}^{k}\varepsilon_{Ut}$
and $\left(T_{2}-2k\right)^{-1/2}\sum_{t=2k+1}^{T_{2}}\varepsilon_{Ut}$.
Therefore, the cross term 
\begin{eqnarray*}
A_{7} & = & \sqrt{\frac{k}{T_{2}}}\left|E\left[\frac{1}{\sqrt{k}}\sum_{t=1}^{k}\varepsilon_{Ut}\frac{1}{\sqrt{T_{2}}}\sum_{t=2k+1}^{T_{2}}\varepsilon_{Ut}\right]\right|=O\left(\sqrt{k/T_{2}}\right)
\end{eqnarray*}
as $N\to\infty$. Our derivations conclude $|\rho_{U}^{*2}-\rho_{U}^{\left(k\right)*2}|=O(\sqrt{k/T_{2}})$
and thus $|\mathcal{Z}_{U}^{\left(k\right)*}-\mathcal{Z}_{U}^{*}|=O_{p}(\sqrt{k/T_{2}})=o_{p}\left(1\right)$
given that $k/T_{2}\to0$ in the condition. This stochastic orders
hold for all $M\in\mathcal{M}$ uniformly and therefore part (b) follows.
\end{proof}
\medskip

In view of Lemma \ref{lem:pop=000026sam} and Lemma \ref{lem:converge_z},
the proof of Theorem \ref{thm:uniform_inference} is an application
of a Berry-Esseen bound for time series. Many results in the probability
theory literature are about strictly stationary time series \citep{bentkus1997berry,jirak2016berry},
but much fewer for heterogeneous time series. The following (\ref{eq:Berry})
comes from \citet{sunklodas1984rate}, which was originally written
in Russian and later was re-interpreted in English in \citet[p.133--134]{sunklodas2000approximation}
and \citet[p.380]{hormann2009berry}.

Define $\alpha_{N}\left(k\right):=\sup_{t\in\mathbb{Z}}\left\{ \left|\Pr\left(AB\right)-\Pr\left(A\right)\Pr\left(B\right)\right|:A\in\mathcal{F}_{N}^{-\infty,t},\ B\in\mathcal{F}_{N}^{t+k,\infty}\right\} $
for $k\in\mathbb{N}$ as the $\alpha$-mixing (strong mixing) coefficient
\citep[p.209]{davidson1994stochastic}. Since the uniform mixing coefficient
$\phi_{N}\left(k\right)\geq\alpha_{N}\left(k\right)$ for each $k$
and $N$, Assumption \ref{assu:alpha_phi_coef} implies $\alpha$-mixing
with a geometric decay rate. For a generic zero-mean time series $\left(x_{t}\right)_{t=1}^{n}$,
if it is $\alpha$-mixing with a geometric decay rate and satisfies
$\max_{t\leq n}\left|x_{t}\right|^{3}\leq\bar{C}<\infty$ and $b_{n}^{2}:=E[(\sum_{t=1}^{n}x_{t})^{2}]\geq n\underline{c}$
for some $\underline{c}>0$ for all $n$ sufficiently large, then
\begin{equation}
\sup_{a\in\mathbb{R}}\left|\Pr\left(\frac{\sum_{t=1}^{n}x_{t}}{b_{n}}\leq a\right)-\Phi\left(a\right)\right|\leq C_{\mathrm{BE}}\frac{\log^{2}b_{n}}{b_{n}}\max_{1\leq t\leq n}E[|x_{t}|^{3}],\label{eq:Berry}
\end{equation}
where $C_{\mathrm{BE}}$ is a constant only depends on the geometric
rate index $\left(c_{1},c_{2}\right)$, $\bar{C}$ and $\underline{c}$.
Thus $C_{\mathrm{BE}}$ is independent of the sample size.
\begin{proof}[Proof of Theorem \ref{thm:uniform_inference}]
 The nominator of the $t$-statistic $\mathcal{Z}_{U}^{*}$ is $T_{2}^{-1/2}\sum_{t\in\mathcal{T}_{2}}\varepsilon_{Ut}$.
Assumption \ref{assu:inference}(c) restricts the third absolute moment
of the summand to be $\max_{t\in\mathcal{T}_{2}}E\left[\left|\varepsilon_{Ut}\right|^{3}\right]\le\max_{t\in\mathcal{T}_{2}}E\left[\left|y_{0t}\right|^{3}\right]\leq\bar{C}$
for some universal constant $\bar{C}$. Under Assumption \ref{assu:inference}(d)
and (e) which regularize the long-run variance, the Berry-Essen bound
(\ref{eq:Berry}) indicates that there exists a constant $C_{\mathrm{BE}}$
such that 
\[
\sup_{a\in\mathbb{R}}\left|\Pr\left(\mathcal{Z}_{U}^{*}\leq a\right)-\Phi\left(a\right)\right|\leq C_{\mathrm{BE}}\frac{\log^{2}\left(\sqrt{T_{2}\rho_{U}^{*2}}\right)}{\sqrt{T_{2}\rho_{U}^{*2}}}\max_{t\in\mathcal{T}_{2}}E\left[\left|\varepsilon_{Ut}\right|^{3}\right].
\]
Notice that in (\ref{eq:Berry}) the universal constant $C_{\mathrm{BE}}$
depends only on $c_{1}$, $c_{2}$, $\bar{C}$ and $\underline{c}$,
we have the uniform rate 
\[
\sup_{M\in\mathcal{M}}\sup_{U\in\mathcal{U}_{R}}\sup_{a\in\mathbb{R}}\left|\Pr\left(\mathcal{Z}_{U}^{*}\leq a\right)-\Phi\left(a\right)\right|=O\left(\sqrt{T_{2}^{-1}\log^{4}T_{2}}\right),
\]
which characterizes the discrepancy between $\Pr\left(\mathcal{Z}_{U}^{*}\leq a\right)$
and the target $\Phi\left(a\right)$. The asymptotic equivalence between
$\mathcal{Z}_{U}^{\left(k\right)*}$ and $\mathcal{Z}_{U}^{*}$ shown
in Lemma \ref{lem:converge_z}(b) implies 
\[
\sup_{M\in\mathcal{M}}\sup_{U\in\mathcal{U}_{R}}\sup_{a\in\mathbb{R}}\left|\Pr\left(\mathcal{Z}_{U}^{\left(k\right)*}\leq a\right)-\Phi\left(a\right)\right|\to0
\]
as $k\to\infty$ and $k/T_{2}\to0$.

\medskip For the generic estimated index set $\check{U}_{R}\subset\mathcal{N}$,
let $\mathcal{U}_{R}^{+}:=\left\{ U\in\mathcal{U}_{R}:\Pr\left(U=\check{U}_{R}\right)>0\right\} $.
The $t$-statistic evaluated on $\check{U}_{R}$ can be explicitly
written as 
\[
\mathcal{Z}_{\check{U}_{R}}=\sum_{U\in\mathcal{U}_{R}}\mathcal{Z}_{U}\boldsymbol{1}\left\{ U=\check{U}_{R}\right\} .
\]
Given the above representation of $\mathcal{Z}_{\check{U}_{R}}$ as
a linear combination of $\left(\mathcal{Z}_{U}\right)_{U\in\mathcal{U}_{R}}$,
its distribution can be characterized as 
\begin{align}
\Pr(\mathcal{Z}_{\check{U}_{R}}\leq a) & =\sum_{U\in\mathcal{U}_{R}}\Pr\left(\mathcal{Z}_{U}\boldsymbol{1}\left\{ U=\check{U}_{R}\right\} \leq a\right)=\sum_{U\in\mathcal{U}_{R}^{+}}\Pr\left(\mathcal{Z}_{U}\boldsymbol{1}\left\{ U=\check{U}_{R}\right\} \leq a\right)\nonumber \\
 & =\sum_{U\in\mathcal{U}_{R}^{+}}\Pr\left(\mathcal{Z}_{U}\leq a|U=\check{U}_{R}\right)\Pr\left(\check{U}_{R}=U\right)\leq\max_{U\in\mathcal{U}_{R}^{+}}\Pr\left(\mathcal{Z}_{U}\leq a|U=\check{U}_{R}\right)\nonumber \\
 & \leq\max_{U\in\mathcal{U}_{R}^{+}}\Pr\left(\mathcal{Z}_{U}^{\left(k\right)*}\leq a+\delta_{a}|U=\check{U}_{R}\right)\label{eq:thm2_1}
\end{align}
where the first equality holds as the events $\left\{ U=\check{U}_{R}\right\} $
are disjoint for those $U\in\mathcal{U}_{R}$, and the last inequality
holds for an arbitrarily small fixed constant $\delta_{a}>0$ when
$N$ is sufficiently large due to the asymptotic equivalence between
$\mathcal{Z}_{\check{U}_{R}}$ and $\mathcal{Z}_{U}^{\left(k\right)*}$
in Lemma \ref{lem:converge_z}(b).

By the definition of the $\phi$-mixing coefficient, Assumption \ref{assu:alpha_phi_coef}
bounds
\begin{equation}
\max_{U\in\mathcal{U}_{R}^{+}}\left|\Pr\left(\mathcal{Z}_{U}^{\left(k\right)*}\leq a+\delta_{a}|U=\check{U}_{R}\right)-\Pr\left(\mathcal{Z}_{U}^{\left(k\right)*}\leq a+\delta_{a}\right)\right|\leq\phi_{N}\left(k\right)\to0\label{eq:thm2_2}
\end{equation}
 as $k\to\infty$ when $N\to\infty$, because $\check{U}_{R}$ is
estimated from the pre-treatment data only so $\left\{ U=\check{U}_{R}\right\} $
is an event in $\mathcal{F}_{-\infty}^{-1}$. We thus continue (\ref{eq:thm2_1}):
\begin{align}
\Pr(\mathcal{Z}_{\check{U}_{R}}\leq a) & \leq\max_{U\in\mathcal{U}_{R}^{+}}\Pr\left(\mathcal{Z}_{U}^{\left(k\right)*}\leq a+\delta_{a}\right)+\phi_{N}\left(k\right)\leq\max_{U\in\mathcal{U}_{R}}\Pr\left(\mathcal{Z}_{U}^{\left(k\right)*}\leq a+\delta_{a}\right)+\phi_{N}\left(k\right)\nonumber \\
 & \leq\max_{U\in\mathcal{U}_{R}}\Pr\left(\mathcal{Z}_{U}\leq a+2\delta_{a}\right)+\phi_{N}\left(k\right)\to\Phi\left(a+2\delta_{a}\right).\label{eq:thm2_3}
\end{align}
where the first inequality follows by applying the triangle inequality
to (\ref{eq:thm2_2}), and the last inequality again due to the asymptotic
equivalence between $\mathcal{Z}_{U}$ and $\mathcal{Z}_{U}^{\left(k\right)*}$
in Lemma \ref{lem:converge_z}. 

Parallel calculation shows that for $N$ sufficiently large we obtain
the lower bound 
\begin{align}
\Pr(\mathcal{Z}_{\check{U}_{R}}\leq a) & =\sum_{U\in\mathcal{U}_{R}^{+}}\Pr\left(\mathcal{Z}_{U}\leq a|U=\check{U}_{R}\right)\Pr\left(U=\check{U}_{R}\right)\nonumber \\
 & \geq\min_{U\in\mathcal{U}_{R}^{+}}\Pr\left(\mathcal{Z}_{U}\leq a|U=\check{U}_{R}\right)\nonumber \\
 & \geq\min_{U\in\mathcal{U}_{R}^{+}}\Pr\left(\mathcal{Z}_{U}^{\left(k\right)*}\leq a-\delta_{a}|U=\check{U}_{R}\right)\nonumber \\
 & \geq\min_{U\in\mathcal{U}_{R}^{+}}\Pr\left(\mathcal{Z}_{U}^{\left(k\right)*}\leq a-\delta_{a}\right)-\phi_{N}\left(k\right)\nonumber \\
 & \geq\min_{U\in\mathcal{U}_{R}}\Pr\left(\mathcal{Z}_{U}\leq a-2\delta_{a}\right)-\phi_{N}\left(k\right)\to\Phi\left(a-2\delta_{a}\right).\label{eq:thm2_4}
\end{align}
The upper bound (\ref{eq:thm2_3}) and the lower bound (\ref{eq:thm2_4})
together restrict $\Pr(\mathcal{Z}_{\check{U}_{R}}\leq a)$ into an
interval 
\[
\Pr(\mathcal{Z}_{\check{U}_{R}}\leq a)\in\left[\Phi\left(a-2\delta_{a}\right),\Phi\left(a+2\delta_{a}\right)\right]
\]
 as $N\to\infty$. Since $\delta_{a}>0$ can be arbitrarily small,
we have $\left|\Pr\left(\mathcal{Z}_{\check{U}_{R}}\leq a\right)-\Phi\left(a\right)\right|\to0$.
Since all the above probability calculations hold uniformly for $M\in\mathcal{M}$
given the definition of $\mathcal{M}$, the conclusion follows.
\end{proof}

\subsection{Proof of Theorem \ref{thm:FS_var_min}}

The following Lemma \ref{lem:inequality} shows the progress that
the greedy algorithm can make. Let $v=\left|V\right|$ and $u=\left|U\right|$
for two generic index sets $V,U\subset\mathcal{N}$. Define $\sigma_{U|V}^{2}:=L_{V}-L_{U}=\sigma_{V}^{2}-\sigma_{U}^{2}$.
\begin{lem}
\label{lem:inequality} Under Assumption \ref{assu:2nd-moment}, for
any set $U,V\subset\mathcal{N}$ such that $U\supset V$ and $u>v$,
we have
\begin{equation}
\max_{j\in\mathcal{N}}\sigma^{2}{}_{\left\{ V,j\right\} |V}\geq\frac{\eta_{u}}{u-v}\,\sigma_{U|V}^{2},\label{eq:key}
\end{equation}
\end{lem}
\begin{rem}
The left-hand side of (\ref{eq:key}) is the magnitude of the descent
of forward selection, and the right-hand side is the proportion of
the total gap $L_{V}$ and $L_{U}$. It means that each greedy pursuit
can narrow the gap $\sigma_{U|V}^{2}$ by a nontrivial proportion.
\end{rem}
\begin{proof}[Proof of Lemma \ref{lem:inequality}]
 In population linear regression models, \citet{das2011submodular}'s
Definition 2.3 defines the \emph{submodularity ratio} $\gamma_{V,k}$,
using our notation, as
\[
\gamma_{V,k}:=\min_{\tilde{V}\subset V,\,\left|S\right|\leq k,\,S\cap L=\emptyset}\frac{\sum_{j\in S}\sigma^{2}{}_{\{\tilde{V},j\}|\tilde{V}}}{\sigma_{(S\cup\tilde{V})|\tilde{V}}^{2}}.
\]
 \citet{das2011submodular}'s Lemma 2.4 states that $\gamma_{V,k}\geq\eta_{v+k}$.
Recall that $\eta_{v+k}$ is our notation for the regularized minimum
eigenvalue defined right above Assumption \ref{assu:eigen}. Let $k=u-v$,
and fix $\tilde{V}=V$ and $S=U\backslash V$. It immediately follows
that 
\begin{align*}
\eta_{u} & \leq\gamma_{V,u-v}\leq\frac{\sum_{j\in U\backslash V}\sigma^{2}{}_{\{V,j\}|V}}{\sigma_{U|V}^{2}}\leq\frac{u-v}{\sigma_{U|V}^{2}}\max_{j\in U}\sigma^{2}{}_{\{V,j\}|V}\leq\frac{u-v}{\sigma_{U|V}^{2}}\max_{j\in\mathcal{N}}\sigma^{2}{}_{\{V,j\}|V}.
\end{align*}
 The stated conclusion follows by rearranging the above inequality.
\end{proof}
We proceed our analysis of forward selection in population. Define
a collection of sequences of index sets
\[
\mathbb{U}_{R}\left(\kappa\right):=\left\{ \left(U_{1},U_{2},\ldots,U_{R}\right)\in\mathcal{N}^{R}\Bigg|\begin{array}{c}
U_{r-1}\subset U_{r},\ \left|U_{r}\backslash U_{r-1}\right|=1,\text{ and }\\
\sigma_{U_{r}|U_{r-1}}^{2}\geq\left(1-\kappa\right)\max_{j\in\mathcal{N}}\sigma_{\{U_{r-1},j\}|U_{r-1}}^{2}
\end{array}\right\} 
\]
for some fixed $\kappa\in\left(0,1\right)$. Any increasing sequence
in $\mathbb{U}_{R}\left(\kappa\right)$ satisfies the inequality $\sigma_{U_{r}|U_{r-1}}^{2}\geq\left(1-\kappa\right)\max_{j\in\mathcal{N}}\sigma_{\left\{ U_{r-1},j\right\} |U_{r-1}}^{2}$.
The constant $\kappa$ can be viewed as a tolerance. We do not have
to be utterly greedy in the sense of capturing the best choice given
$U_{r-1}$. As long as we make progress in each iteration by reducing
the gap to at least a constant proportion of what the most greedy
choice can achieve, we can still approach, or even surpass, our target.
This is the message of the following lemma.
\begin{lem}
\label{lem:population} For any sequence of sets $\left(U_{1},\ldots,U_{R}\right)\in\mathbb{U}_{R}\left(\kappa\right)$
and any $W\subset\mathcal{N}$, we have 
\begin{equation}
\sigma_{U_{R}}^{2}-\sigma_{W}^{2}\leq\sigma_{\emptyset}^{2}\left(1-\left(1-\kappa\right)\eta_{w+R}/w\right)^{R}\label{eq:diff_UR_W}
\end{equation}
where $w=\left|W\right|.$
\end{lem}
\begin{rem}
Lemma \ref{lem:population} states what happens when the forward selection
algorithm is applied to the population model. In each iteration, the
index set includes one more element; however the variance updates
less greedily. Even with this less greedy algorithm, given the optimal
set $W=U_{u}^{*}$, after $R$-times iteration with $R=R\left(N\right)\to\infty$
as $N\to\infty$, the difference between $\sigma_{U_{R}}^{2}$ and
the optimal $\sigma_{U_{u}^{*}}^{2}$ will shrink to zero. The tolerance
$\kappa$ will be needed when we bring the population greedy algorithm
to the data where sampling errors must be accommodated. This inequality
(\ref{eq:diff_UR_W}) holds trivially when the left-hand side takes
negative values.
\end{rem}
\begin{proof}[Proof of Lemma \ref{lem:population}]
 We first derive an inequality for generic sets $W,V\subset\mathcal{N}$
and $W\neq V$. Define in this proof $U=W\cup V$ so that $U\supset V$
and $u-v\geq1$. Since $u-v=\left|W\cup V\right|-v\leq w$, the restricted
minimum eigenvalues satisfies $\eta_{u}=\eta_{\left|W\cup V\right|}\geq\eta_{w+v}$
and it implies\textbf{
\[
\frac{\eta_{u}}{u-v}\sigma_{U|V}^{2}\geq\frac{\eta_{w+v}}{w}\sigma_{U|V}^{2}=\frac{\eta_{w+v}}{w}\left(\sigma_{V}^{2}-\sigma_{U}^{2}\right)\geq\frac{\eta_{w+v}}{w}\left(\sigma_{V}^{2}-\sigma_{W}^{2}\right),
\]
}where the last inequality follows as $\sigma_{U}^{2}\le\sigma_{W}^{2}$.
Multiply $-\left(1-\kappa\right)$ and add $\left(\sigma_{V}^{2}-\sigma_{W}^{2}\right)$
on both sides of the above inequality:
\begin{align}
\left(1-\left(1-\kappa\right)\frac{\eta_{w+v}}{w}\right)\left(\sigma_{V}^{2}-\sigma_{W}^{2}\right) & \geq\left(\sigma_{V}^{2}-\sigma_{W}^{2}\right)-\left(1-\kappa\right)\frac{\eta_{u}}{u-v}\sigma_{U|V}^{2}\nonumber \\
 & \geq\left(\sigma_{V}^{2}-\sigma_{W}^{2}\right)-\left(1-\kappa\right)\max_{j\in\mathcal{N}}\sigma^{2}{}_{\left\{ V,j\right\} |V}\label{eq:b_uv}
\end{align}
where the second inequality follows by Lemma \ref{lem:inequality}.

Now we substitute the generic $V$ with the specific choice $U_{R}$.

Case (i): If $\sigma_{U_{R}}^{2}<\sigma_{W}^{2},$ then (\ref{eq:diff_UR_W})
holds trivially. 

Case (ii): If $\sigma_{U_{R}}^{2}\geq\sigma_{W}^{2}$, then 
\begin{align}
0 & \leq\sigma_{U_{R}}^{2}-\sigma_{W}^{2}=(\sigma_{U_{R-1}}^{2}-\sigma_{W}^{2})-\sigma_{U_{R}|U_{R-1}}^{2}\leq(\sigma_{U_{R-1}}^{2}-\sigma_{W}^{2})-\left(1-\kappa\right)\max_{j\in\mathcal{N}}\sigma_{\left\{ U_{R-1},j\right\} |U_{R-1}}^{2}\nonumber \\
 & \leq\left(1-\left(1-\kappa\right)\eta_{w+R}/w\right)(\sigma_{U_{R-1}}^{2}-\sigma_{W}^{2}),\label{eq:R-1}
\end{align}
where the second inequality holds by the definition of $\mathbb{U}_{R}\left(\kappa\right)$,
and the third inequality by (\ref{eq:b_uv}). The fact that $\sigma_{U_{r}}^{2}$
is (weakly) monotonically decreasing in $r$ implies 
\begin{equation}
0\leq\sigma_{U_{R-1}}^{2}-\sigma_{W}^{2}\leq\left(1-\left(1-\kappa\right)\eta_{w+R-1}/w\right)(\sigma_{U_{R-2}}^{2}-\sigma_{W}^{2})\label{eq:R-2}
\end{equation}
and more generally 
\begin{equation}
0\leq\sigma_{U_{r}}^{2}-\sigma_{W}^{2}\leq\left(1-\left(1-\kappa\right)\eta_{w+r}/w\right)\left(\sigma_{U_{r-1}}^{2}-\sigma_{W}^{2}\right),\ \ \text{for all }2\leq r\leq R.\label{eq:R-3}
\end{equation}
Substitute (\ref{eq:R-2}) into (\ref{eq:R-1}), and iterate the inequality
(\ref{eq:R-3}):
\begin{align*}
\sigma_{U_{R}}^{2}-\sigma_{W}^{2} & \leq\left(1-\left(1-\kappa\right)\frac{\eta_{w+R}}{w}\right)\left(1-\left(1-\kappa\right)\frac{\eta_{w+R-1}}{w}\right)(\sigma_{U_{R-2}}^{2}-\sigma_{W}^{2})\\
 & \leq\cdots\leq\left(\sigma_{U_{1}}^{2}-\sigma_{W}^{2}\right)\prod_{r=1}^{R}\left(1-\left(1-\kappa\right)\eta_{w+r}/w\right)\leq\sigma_{\emptyset}^{2}\left(1-\left(1-\kappa\right)\eta_{w+R}/w\right)^{R},
\end{align*}
where the last inequality holds as $\sigma_{U_{1}}^{2}-\sigma_{W}^{2}\le\sigma_{\emptyset}^{2}-\sigma_{W}^{2}\leq\sigma_{\emptyset}^{2}$
and $\eta_{W+r}$ is (weakly) monotonically decreasing in $r\leq R$.
\end{proof}
The calculations in Lemmas \ref{lem:inequality} and \ref{lem:population}
are carried out in the population regression model. Next, we link
the population model to the sample to prove Theorem \ref{thm:FS_var_min}.
\begin{proof}[Proof of Theorem \ref{thm:FS_var_min}]
 By adding and subtracting, 
\begin{equation}
\hat{\sigma}_{\hat{U}_{R}}^{2}-\sigma_{U_{u}^{*}}^{2}=(\hat{\sigma}_{\hat{U}_{R}}^{2}-\sigma_{\hat{U}_{R}}^{2})+(\sigma_{\hat{U}_{R}}^{2}-\sigma_{U_{u}^{*}}^{2})\label{eq:3terms}
\end{equation}
is decomposed into two terms. Since $|\hat{U}_{R}|=R$, we invoke
Lemma \ref{lem:pop=000026sam} so that 
\begin{equation}
\hat{\sigma}_{\hat{U}_{R}}^{2}-\sigma_{\hat{U}_{R}}^{2}=O_{p}(\sqrt{R\left(\log N\right)/T_{1}})=o_{p}(1).\label{eq:sigma_app}
\end{equation}
 We focus on the second term $\sigma_{\hat{U}_{R}}^{2}-\sigma_{U_{u}^{*}}^{2}$
in (\ref{eq:3terms}). Let $\varrho_{r}:=\max_{\mathcal{U}_{r}}\left|\widehat{\sigma}_{U}^{2}-\sigma_{U}^{2}\right|.$
Define a collection of sets
\begin{equation}
\mathcal{A}_{r}\left(\kappa\right):=\big\{ V\subset\mathcal{N}:\left|V\right|=r,\ \max_{j\in\mathcal{N}}\sigma_{\left\{ j,V\right\} |V}^{2}>4\varrho_{r}/\kappa\big\}.\label{eq:event}
\end{equation}
Let $\tilde{j}=\tilde{j}\left(V\right):=\arg\text{\ensuremath{\max}}_{j\in\mathcal{N}}\,\widehat{\sigma}_{\left\{ j,V\right\} |V}^{2}$,
which is the index selected by the greedy algorithm from the sample
given the set $V$. Denote $(\widehat{U}_{1},\ldots,\widehat{U}_{R})$
as the selected sequence by the greedy algorithm. We discuss two cases.

Case (i): If $\widehat{U}_{r}\in\mathcal{A}_{r}\left(\kappa\right)$
for all $2\leq r\leq R$, then 
\begin{eqnarray*}
\sigma_{\left\{ \tilde{j},\widehat{U}_{r-1}\right\} |\widehat{U}_{r-1}}^{2} & \geq & \widehat{\sigma}_{\{\tilde{j},\widehat{U}_{r-1}\}|\widehat{U}_{r-1}}^{2}-\big|\widehat{\sigma}_{\{\tilde{j},\widehat{U}_{r-1}\}|\widehat{U}_{r-1}}^{2}-\sigma_{\{\tilde{j},\widehat{U}_{r-1}\}|\widehat{U}_{r-1}}^{2}\big|\\
 & \geq & \widehat{\sigma}_{\{\tilde{j},\widehat{U}_{r-1}\}|\widehat{U}_{r-1}}^{2}-2\max_{\mathcal{U}_{r}}\left|\widehat{\sigma}_{U}^{2}-\sigma_{U}^{2}\right|=\widehat{\sigma}_{\{\tilde{j},\widehat{U}_{r-1}\}|\widehat{U}_{r-1}}^{2}-2\varrho_{r-1}\\
 & = & \max_{j\in\mathcal{N}}\widehat{\sigma}_{\{j,\widehat{U}_{r-1}\}|\widehat{U}_{r-1}}^{2}-2\varrho_{r-1}\\
 & \geq & \max_{j\in\mathcal{N}}\left\{ \sigma_{\{j,\widehat{U}_{r-1}\}|\widehat{U}_{r-1}}^{2}-\big|\widehat{\sigma}_{\{j,\widehat{U}_{r-1}\}|\widehat{U}_{r-1}}^{2}-\sigma_{\{j,\widehat{U}_{r-1}\}|\widehat{U}_{r-1}}^{2}\big|\right\} -2\varrho_{r-1}\\
 & \geq & \max_{j\in\mathcal{N}}\sigma_{\{j,\widehat{U}_{r-1}\}|\widehat{U}_{r-1}}^{2}-4\varrho_{r-1}\\
 & > & (1-\kappa)\max_{j\in\mathcal{N}}\sigma_{\{j,\widehat{U}_{r-1}\}|\widehat{U}_{r-1}}^{2},
\end{eqnarray*}
where the second and the fourth inequalities follow by adding and
subtracting as in (\ref{eq:3terms}). Thus we have $(\widehat{U}_{1},\ldots,\widehat{U}_{R})\in\mathbb{U}_{R}\left(\kappa\right)$.
When $W=U_{u}^{*}$, by Assumption \ref{assu:eigen} and Lemma \ref{lem:population}
we have 
\begin{equation}
\sigma_{\widehat{U}_{R}}^{2}-\sigma_{u}^{*2}\leq\sigma_{\emptyset}^{2}\left(1-\left(1-\kappa\right)\eta_{u+R}/u\right)^{R}\leq\sigma_{\emptyset}^{2}\left(1-\left(1-\kappa\right)c/u\right)^{R}\to0\label{eq:conc1}
\end{equation}
when the event $\left\{ (\widehat{U}_{1},\ldots,\widehat{U}_{R})\in\mathbb{U}_{R}\left(\kappa\right)\right\} $
occurs, and the limit holds by $u/R\to0$ as $N\to\infty$.

Case (ii): Suppose the selected sequence $(\widehat{U}_{1},\ldots,\widehat{U}_{R})$
has some elements $\widehat{U}_{r}$ not satisfying $\mathcal{A}_{r}\left(\kappa\right)$.
Let $\tilde{r}:=\min\{r\in\left\{ 1,\ldots,R\right\} |\widehat{U}_{r}\notin\mathcal{A}_{r}\left(\kappa\right)\}$
be the first occurrence of violation when the sequence of selection
progresses, and by the definition of $\mathcal{A}_{r}\left(\kappa\right)$
we have 
\begin{equation}
\max_{j\in\mathcal{N}}\sigma_{\{j,\widehat{U}_{\tilde{r}}\}|\widehat{U}_{\tilde{r}}}^{2}\leq4\varrho_{r}/\kappa.\label{eq:small_case}
\end{equation}
 If $U_{u}^{*}\subset\widehat{U}_{\tilde{r}}$, which is the ideal
case when the selected set includes the population optimal set $U_{u}^{*}$,
then $\sigma_{\widehat{U}_{R}}^{2}\leq\sigma_{\widehat{U}_{\tilde{r}}}^{2}\leq\sigma_{U_{u}^{*}}^{2}$.
On the other hand, even if $U_{u}^{*}$ is not a subset of $\widehat{U}_{\tilde{r}}$,
we have
\begin{align}
\sigma_{\widehat{U}_{R}}^{2}-\sigma_{u}^{*2} & \leq\sigma_{\widehat{U}_{\tilde{r}}}^{2}-\sigma_{u}^{*2}\leq\sigma_{\widehat{U}_{\tilde{r}}}^{2}-\sigma_{U_{u}^{*}\cup\widehat{U}_{\tilde{r}}}^{2}=\sigma_{(U_{u}^{*}\cup\widehat{U}_{\tilde{r}})|\widehat{U}_{\tilde{r}}}^{2}\leq\frac{u}{\eta_{u+\tilde{r}}}\cdot\max_{j\in\mathcal{N}}\sigma_{\{j,\widehat{U}_{\tilde{r}}\}|\widehat{U}_{\tilde{r}}}^{2}\nonumber \\
 & \leq\frac{u}{\eta_{u+R}}\cdot\max_{j\in\mathcal{N}}\sigma_{\{j,\widehat{U}_{\tilde{r}}\}|\widehat{U}_{\tilde{r}}}^{2}\leq\frac{u}{c}\cdot\frac{4\varrho_{\tilde{r}}}{\kappa}=o_{p}\left(\sqrt{R^{3}\left(\log N\right)/T_{1}}\right),\label{eq:conc2}
\end{align}
where the third inequality follows by Lemma \ref{lem:inequality},
the fifth inequality by the condition (\ref{eq:small_case}) and Assumption
\ref{assu:eigen} since $u+R\leq\left(1+\delta_{1}\right)R$ holds
asymptotically for any $\delta_{1}>0$ as $u/R\to0$ when $N\to\infty$,
and finally the stochastic order of $\varrho_{\tilde{r}}$ by Lemma
\ref{lem:pop=000026sam}. The statement of the theorem follows by
collecting collecting (\ref{eq:conc1}), (\ref{eq:conc2}) and (\ref{eq:sigma_app})
and substituting them into (\ref{eq:3terms}).
\end{proof}
\bigskip 
%\small
%\singlespacing

\bibliographystyle{elsarticle-harv}
\bibliography{corruption}

\end{document}